\documentclass[%
 reprint,
 amsmath,amssymb,
 aps,
pra,
]{revtex4-1}
\usepackage{hyperref}                         % %
\usepackage[all]{hypcap}                      % %
\usepackage{graphicx}% Include figure files
\usepackage{dcolumn}% Align table columns on decimal point
\usepackage{bm}% bold math
\usepackage{xcolor}
\usepackage{float}
\begin{document}

\preprint{APS/123-QED}

\title{Bose-Fermi trasmutation for one-dimensional
	harmonic trap}% Force line breaks with \\

%\author{R. Avella}
%\author{J. J. Mendoza-Arenas}
%\author{R. Franco}
%\author{J. Silva-Valencia.}

\noindent
\author{J. Nisperuza $^1$, JP Rubio$^1$ and R. Avella$^1$}% \author{J. J. Mendoza-Arenas$^2$, R. Franco$^1$ and J. Silva-Valencia$^1$}\\
\altaffiliation[]{rgavellas@libertadores.edu.co}%Lines break automatically or can be forced with \\
\affiliation{%
$^1$ Facultad de Ingenieria Aeronautica, Fundaci\'{o}n Universitaria los Libertadores, A. A. 75087 Bogot\'{a}, Colombia.
%$^2$Departamento de F\'{\i}sica, Universidad de los Andes, A. A. 4976 Bogot\'{a}, Colombia %\textbackslash\textbackslash
}%

%\author{J. J. Mendoza-Arenas}
%\affiliation{
%\affiliation{
%\author{R. Franco}
%\affiliation{%
%\author{R. Franco}
%\affiliation{%
%	Authors' institution and/or address\\
%	This line break forced with \textbackslash\textbackslash
%\author{J. Silva-Valencia}
%\affiliation{%
%	Authors' institution and/or address\\
%	This line break forced with \textbackslash\textbackslash

\date{\today}% It is always \today, today,

\begin{abstract}
Using Density Matrix renormalization group (DMRG), we study the ground state properties of spin one-half fermions and scalar bosons in the soft-core limit, with weak s-wave inter and intra species interactions. We considered the system subject to one-dimensional (1D) optical lattice and a superimposed potential at zero temperature, in the framework of  Bose-Fermi-Hubbard model. We found that for certain fillings and interaction parameters, a transmutation occurs between the ground states of bosons and fermions when the densities are exchanged. We too report that the density distributions of bosons and fermions overlap with each other in a bosonic and fermionic Mott plateau, when the interaction parameters fulfill the relationship $U_{BB}>U_{BF}<U_{BF}$. We also find that the fermions are repelled out of the central region of the trap for sufficiently strong $U_{FF}$ interaction, exhibiting phase separation of Bose and Fermi components.

%\begin{description}
%\end{description}
\end{abstract}

\maketitle

%\tableofcontents

\section{INTRODUCTION}

In the last years much progress has been achieved in the development of laser cooling \cite{Kerman00} and optical trapping technology\cite{Greiner01,Greiner02, Orzel01}. These progress has opened the way to load within external potentials, mixtures of ultracold atomic gases with different statistics and allow to have complete control over all physical parameters \cite{Anglin02, Ospelkaus06, Hadzibabic03, Inouye04}. Due to such high controllability, these systems have become an important practical tool for applications in quantum control, information processing \cite{ Jessen01, Duan03} , quantum computation \cite{Deutsch00, Jaksch99,Garcia03, Dorner03, Pachos03}, besides the interest in the investigation of quantum phenomena \cite{Sachdev99, Jaksch98, van01, Ruostekoski02, Hofstetter02, Paredes03, Recati03, Buchler03}. These investigations have opened the possibility of studying experimentally the influence of interactions between species, for example, the pairing of fermions that is analogous to the formation of Cooper pairs in the BCS model \cite{Bijlsma00, Heiselberg00, Viverit02} and the arise of different quantum phases \cite{Pollet08, McNamara06, Klempt08, Karpiuk06, Schneider08, Lous18, Wille06, Sugawa11, Akdeniz02, Vichi98}.

Theoretically, numerous studies have been carried out to analyze the Bose-Fermi mixture (BFM) \cite{Roth02, Liu03, Modugno03, Adhikari04, Miyakawa01, Karpiuk05, Pelster07} indicating for example an asymmetry between the attraction and repulsion cases \cite{Best09, Albus03}, as well as phase separation, spatial modulation \cite{Polak10}, supersolid phase, charge density wave \cite{Titvinidze08, Avella19} and the boson phase transition from the Mott insulator to super fluid \cite{Mering08, Bukov14, Fehrmann04, Avella20}. 

Intrinsically quantum gases are inhomogeneous due to confinement potential additional external, which is generally roughly harmonic, but it is also possible to design quasi-homogeneous systems \cite{Bergeman03}. One of the simplest confinements that has been extensively studied for the possibility of doing it experimentally corresponds to harmonic traps, because any potential can be approximated to a harmonic oscillator at a local minimum \cite{Laird17} and because as a consequence of confinement, the different phases of many-body systems coexist inside the trap \cite{Sugawa11}.

In conventional condensed-matter physics, Bose-Fermi (BF) mixture realized in a harmonic trap, and in optical lattices are not easy to investigate \cite{Noda2012}. Experimentally the 1D systems can be realized by confining the cold atoms in two perpendicular optical lattices or in strong anisotropic magnetic traps\cite{Moritz04, Wang12, Paredes04, Kinoshita04}. These accomplishments has stimulated theoretical investigation on related topic to the  quasi-1D Bose-Fermi mixture, which might provide theoretical guidance on the potential experimental implementation\cite{Noda2012}. Theoretical investigations have focused on the phase diagrams, ground state and thermodynamical properties in the scheme of Luttinger liquid theory \cite{Cazalilla03, Mathey04} and Bethe ansatz method \cite{Imambekov06, Imambekov06R, Batchelor05}. These approaches have proposed a whole variety of quantum phases present in homogeneous Bose-Fermi mixtures at low temperature\cite{Pollet08}, such as: charge-density wave, fermionic pairing phase, polaronic properties, phase separation and even supersolid state \cite{Titvinidze08, Titvinidze09}. The richer phase structure in comparison with the single component systems, is induced by the competition among inter and intra species interactions\cite{Hao11}.

Inhomogeneous systems have been studied considering mixture of spinless bosons and spin-polarized fermions. These systems bring novel issues, such as the spatial structure of the ground state and  partial demixing of the two clouds \cite{Imambekov06, Imambekov06R, Xiangguo09}. A particularly interesting feature of these mixtures, is the thin line separating the properties of bosons and fermions, for example in the Tonks-Girardeau gas limit of infinitely strong boson-fermion repulsions a ground-state is highly degenerate \cite{Girardeau06} and is associated to the freedom of fixing the sign of many-body wave function under the exchange of a boson with a fermion \cite{Guan09, Wang09, Yang09}. This result is known as the Bose-Fermi mapping theorem for hard core particles \cite{Fang11} and was realized experimentally for firts time in \cite{Paredes04}. However the proof that bosons are dual to non-interacting fermions, was realized for Cheon and Shigehara, considering hard core has zero range and in absence of other interactions \cite{Cheon99}.

The Cheon-Shigehara mapping was used to find a duality relation in systems composed by spin-1/2 fermions interacting with two-component bosons \cite{Girardeau04} and an exact analytical solution of point hard core Bose-Fermi considering spin-1 bosons \cite{PhysRevLett.100.160405}. Recently M. Valiente found the most general one-to-one mapping between quantum particles with different statistics in one spatial dimension with arbitrary low energy interactions. This general mapping not restricted to pairwise forces, is valid for arbitrary single-particle dispersion, including non-relativistic, relativistic, continuum limits of lattice Hamiltonians and can also be applied to any internal structure and spin \cite{Valiente20}.

This leaves open the possibility of studying the transmutation of bosons into fermions and vice versa in a system of spinless bosons in the soft-core limit and spin one-half fermions, that is studied in this paper. Firts we construct an effective single-band Bose-Fermi Hubbard Hamiltonian \cite{Albus03} and use Density Matrix renormalization group (DMRG) \cite{White93} to study the ground state properties of the 1D Bose-Fermi mixtures, subject to an optical lattice and a superimposed potential at zero temperature. 

%%%%%%%%%%%%%%%%%%%%%%%%%%%%%%%%%%%%%%%%%%%%%%%%%%%%%%%%%%%%%%%%%%%%%%%%%%%%%%%%%%%%%%%%%%%%%%%%%%%%%%%%%%%%%
The paper is organized as follows. The model used to describe a mixture of bosonic and fermionic atoms is introduced in Sec. \ref{sec:Bose-Fermi mixtures model}. We explore the different parameters to obtein Bose-Fermi transmutation and characterize the different ground states in Sec. \ref{sec:Bose-Fermi transmutation}. Finally, in Sec. \ref{sec:Conclusions} we establish our final remarks.

\section{Bose-Fermi mixtures model}\label{sec:Bose-Fermi mixtures model}
We study a mixture of $N_B$ scalar bosons in the soft-core limit and $N_F$ spin one-half fermions in one-dimensional optical lattice and a superimposed trapping harmonic potential at zero temperature. Bosons and fermions experience the same external potential, and their masses are equal. This condition can be satisfied for combination of bosonic and fermionic isotopes of the same element, such as $^{170}Yb-^{171}Yb$ and can be described by the Bose-Fermi-Hubbard Hamiltonian

\begin{equation}
\label{hubofear}
\begin{split}
\hat{H}_{BF}=&-t_B\sum_{\langle i,j\rangle}\Big(\hat{b}_{i}^{\dag}\hat{b}_{j}+H.c.\Big)+
\frac{U_{BB}}{2}\sum_{i}\hat{n}_{i}^B(\hat{n}_{i}^B-1)\\
&-t_F\sum_{\langle i,j\rangle\sigma}\Big(\hat{f}_{i,\sigma}^{\dag}\hat{f}_{j,\sigma}+H.c.\Big)
+\frac{U_{FF}}{2}\sum_{i,\sigma\neq\sigma'}\hat{n}_{i,\sigma}^F\hat{n}_{i,\sigma'}^F\\
&+W\sum_{i,\sigma}\Big(i-\frac{L}{2}\Big)^2\Big(\hat{n}_{i}^B+\hat{n}_{i,\sigma}^F\Big)+U_{BF}\sum_{i}\hat{n}_i^B\hat{n}_i^F,
\end{split}
\end{equation}

\begin{figure}[H]
	\centering
	\includegraphics[scale=0.35]{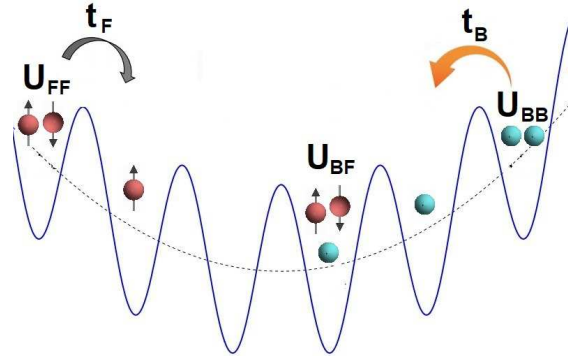}\\
	\caption{{\footnotesize Bose-Fermi-Hubbard-Hamiltonian model, for the system composed by scalar bosons (magenta circles) in the soft-core limit and spin one-half fermions (red circles), in one-dimensional optical lattice and a superimposed trapping harmonic potential (black dotted lines) at zero temperature. The on-site boson-boson, fermion-fermion and boson-fermion interaction parameters are represent by $U_{BB}$, $U_{FF}$ and $U_{BF}$ respectively. $W$ is the amplitude of the trap.}}
	\label{harmonico1}
\end{figure}

Where $\hat{b}^{\dag}_i$ ($\hat{f}^{\dag}_{i,\sigma}$) is the bosonic (fermionic of spin $\sigma=\uparrow,\downarrow$) creation operator at site $i$ of a lattice of size $L$, while $\hat{n}_{i}^B$ ($\hat{n}_{i,\sigma}^F$) denotes the local boson (fermion) number operator. The on-site boson-boson $U_{BB}$ (fermion-fermion $U_{FF}$) and boson-fermion $U_{BF}$ interaction parameters are related to the bosonic (fermoinic) $a_{bb}$ ($a_{ff})$ and the Bose-Fermi $a_{bf}$ scattering length \cite{Albus03} and are s-wave repulsive contact interactions . The index $\langle i,j\rangle$ denotes summation over nearest neighbors and $t_{B(F)}$ is the tunneling amplitude for bosons (fermions). In the following, energies and gaps are measured in units of the fermionic hopping and since the bosons and fermions masses in our system are approximately equal, we have $t_B=t_F= 1$. The harmonic trapping potential is incorporated through the term $W\sum_{i,\sigma}\Big(i-\frac{L}{2}\Big)^2\Big(\hat{n}_{i}^B+\hat{n}_{i,\sigma}^F\Big)$, where $W$ is the amplitude of the trap. A representation of the model is illustrated in Figure \ref{harmonico1}.

In this study we consider a lattice of size $L=60$ sites, due to the experimental possibility of confining quantum gases in harmonic potentials with up to one hundred sites. The number of fermions and bosons particles per site, takes values in the interval [0,2] and [0,3] respectively. The maximum value of $n_{max}^B=3$, is considered because it has been argued in several reports that the qualitative physical properties obtained for $n_{max}^B=3$ do not change when $n_{max}^B$ is increased \cite{Pai96, Rossini12}. 

Due to the inhomogeneous distribution of atoms in the system, we focused on local quantities in order to characterize the ground state of the system, such as the local number of localized bosons (fermions)  $\Big\langle \hat{n}_i^{B(F)}\Big\rangle = \Big\langle\sum_\sigma\hat{n}_{i,\sigma}^{B(F)}\Big\rangle$, the total local number of localized atoms $\langle \hat{n}_i^T  \rangle = \langle\hat{n}_{i}^B+\sum_\sigma\hat{n}_{i,\sigma}^F\rangle$ and the variance of the local total density per site $\Delta \hat{n}_i^{B(F)}=\langle \hat{n}_i^2\rangle-\langle n_i\rangle^2$\cite{Batrouni02, Rigol08}. We perform several finite-system sweeps until the ground-state energy is converged to an absolute error of $10^3$ , keeping a discarded weight of $\approx10^{-7}$ in the dynamic
block selection state (DBSS) protocol \cite{Legeza03}.

\begin{figure*}
	\includegraphics[scale=0.5]{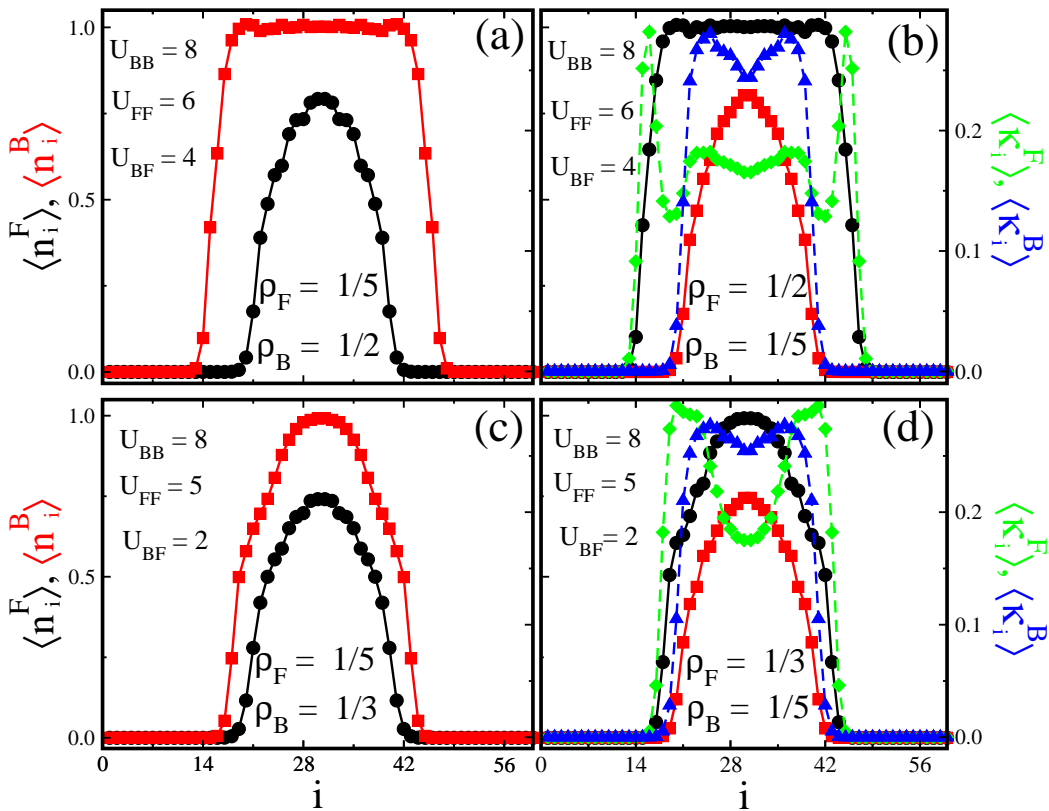}
	\caption{{\footnotesize Bose-Fermi transmutation. The fermions and boson densities are represent in the figure by  black circles and red squares respectively. The variance of the local total fermions and boson densities per site is shown as green diamonds and blue triangles respectively. (a)-(b) Transmutation of bosons into fermions and vice versa for interactions $U_{BB}=8$, $U_{FF}=6$ and $U_{BF}=4$. The transmutation arise when densities are exchanged $\rho_F=1/5\rightarrow\rho_F=1/2$ and $\rho_B=1/2\rightarrow\rho_B=1/5$. (c)-(d) Transmutation of bosons into fermions and vice versa for interactions $U_{BB}=8$, $U_{FF}=5$ and $U_{BF}=2$. Transmutation arise when densities are exchanged $\rho_F=1/5\rightarrow\rho_F=1/3$ and $\rho_B=1/3\rightarrow\rho_B=1/5$.}}
	\label{sim}
\end{figure*}

\section{Bose-Fermi transmutation}\label{sec:Bose-Fermi transmutation}

In this section we study the transmutation of bosons into fermions and vice versa, in a 1D system of $N_B$ spinless bosons and $N_F$ one-half spin fermions that interact in the same harmonic confinement potential. To carry out this study, we consider that boson-boson, fermion-fermion and boson-fermion interactions ($U_{BB}, U_{FF}$ and $U_{BF}$ respectively) are repulsive and amplitude of the trap that confine the both species is W$=0.03$. 

First we fix boson and fermion density number in $\rho_B=1/2$ and $\rho_F=1/5$ respectively, we also fix the fermion-fermion $U_{FF}=6$ and boson-fermion $U_{BF}=4$ interactions and we consider that the boson-boson interaction is variable. In figure \ref{sim}(a) we show the ground state for $U_{BB}=8$.

The ground state of this configuration, presents the coexistence of insulator bosonic and fermionic band states at the ends of trap followed by a phase separation region between the sites 13-19 (42-48) where there are bosons in the superfluid state and there are no fermions. This phase was reported for one-dimensional harmonically trapped Bose-Fermi mixture in \cite{Dehkharghani17, Xianlong13, Albus03, Wang12}. In the center of potential due to the greater interaction between bosons, these tend to occupy each one a site, giving rise to a bosonic Mott insulator state and fermions form a superfluid state due to the low density and the repulsive character of the interactions.

When densities are exchanged i.e., $\rho_F=1/5\rightarrow\rho_F=1/2$ and $\rho_B=1/2\rightarrow\rho_B=1/5$  (figure \ref{sim} (b)) the number occupation of each species tends to be reversed. The ends of trap, present a bosonic (red squares) and fermionic (black circles) occupation number equal to zero $\Big(\Big\langle \hat{n}_i^{B}\Big\rangle=\Big\langle \hat{n}_i^{F}\Big\rangle =0\Big)$, which indicates that both bosons and fermions are in a band insulator state. This phase is characterized by a constant bosonic (blue triangles) and fermionic (green diamonds) compressibility equal to zero ($\kappa_i^{B}=\kappa_i^{F}=0$).

The ground state also exhibits a greater confinement of the bosons, which gives rise to a region between the sites 13-19 (42-48) where there are fermions in the superfluid state and there are no bosons, giving rise to a phase separation. In the center of the trap the bosons are in the superfluid states and the fermions form a Mott insulator state with one fermion per site, characterized by a constant compressibility and different of zero $\kappa_i^{\textbf{B}}\neq0$ (green diamonds). If the figure \ref{sim} (a) is compared with figure \ref{sim} (b), it is observed that a transmutation of the bosons into fermions and vice versa took place as presented in \cite{Valiente20}. This being one of the most important observation in this investigation.

The transmutation is also present in figures \ref{sim}(c) and \ref{sim}(d), considering $U_{BB}=8$, $U_{FF}=5$, $U_{BF}=2$ and the exchange of densities $\rho_F=1/5\rightarrow\rho_F=1/3$ and $\rho_B=1/3\rightarrow\rho_B=1/5$ respectively. Characterization of the ground state is presented in the figure \ref{sim} (d), where $\rho_F=1/3$ and $\rho_B=1/5$. This presents the coexistence of insulator bosonic and fermionic band states at the ends of the trap, with a bosonic (red squares) and fermionic (black circles) occupancy number equal to zero $\Big(\Big\langle \hat{n}_i^{B}\Big\rangle=\Big\langle \hat{n}_i^{F}\Big\rangle =0\Big)$ and characterized by a constant bosonic (blue triangles) and fermionic (green diamonds) compressibility equal to zero ($\kappa_i^{B}=\kappa_i^{F}=0$). The center of trap presents the coexistence of bosonic and fermionic superfluid, characterized by a variable compressibility and with a higher occupation number for fermions in relation with the bosons. This ground state is reversed when the densities are exchange i.e. $\rho_F=1/5$ and $\rho_B=1/3$ as illustrated in the figure \ref{sim} (c).

\begin{figure}[H]
	\centering
	\includegraphics[scale=0.32]{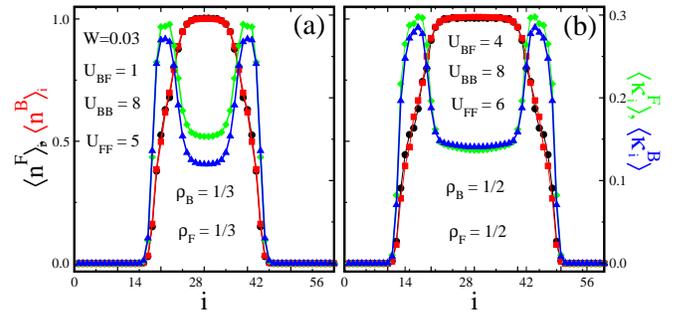}\\
	\caption{{\footnotesize Coexistence of bosonic and fermionic Mott plateau. The fermion (boson) density is represent in the figure by  black circles (red squareds). The variance of the local total fermion (boson) density per site is shown as green diamonds (triangles). (a)   coexistence of bosonic and fermionic Mott plateau for interactions $U_{BB}=8$, $U_{FF}=5$, $U_{BF}=1$ and $\rho_B=\rho_F=1/3$. (b) coexistence of bosonic and fermionic Mott plateau for interactions $U_{BB}=8$, $U_{FF}=6$, $U_{BF}=4$ and $\rho_B=\rho_F=1/2$.}}
	\label{insulator}
\end{figure}

\begin{figure}[H]
	\centering
	\includegraphics[scale=0.325]{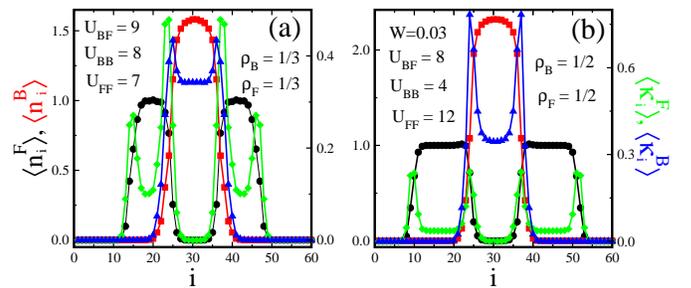}\\
	\caption{{\footnotesize Phase separation. The fermion (boson) density is represent in the figure by  black circles (red squareds). The variance of local total fermion (boson) density per site is shown as green diamonds (blue triangles). (a) Phase separation for interactions $U_{BB}=4$, $U_{FF}=12$, $U_{BF}=6$ and $\rho_B=\rho_F=1/3$. (b) Phase separation for interactions $U_{BB}=4$, $U_{FF}=6$, $U_{BF}=8$ and $\rho_B=\rho_F=1/2$.}}
	\label{separation}
\end{figure}

%\section{Other ground states}
Transmutation is also present when bosons and fermions have the same densities as shown in figures \ref{insulator} and \ref{separation}, considering the effects of on-site interaction parameters. First we discuss the case of $\rho_B=\rho_F=1/3$, $U_{BB}=8, U_{BF}=1$ and $U_{FF}=5$ as shown in the figure \ref{insulator}(a), where the number occupation $\hat{n}_i^B$ (red squares) and $\hat{n}_i^F$ (black circles) are plotted as a function of the site ($i$) and they overlap each other, verifying the existence of Bose-Fermi transmutation. The ground state in the center of  trap is composed by the coexistence of bosonic and fermionic Mott plateau, as verified by boson (green diamonds) and fermion (blue triangles) compressibility, which is constant and different from zero. 

A similar ground states is presented for the case $\rho_B=\rho_F=1/2$, $U_{BB}=8, U_{BF}=4$ and $U_{FF}=6$ (figure \ref{insulator}(b)), where the relationship $\rho_B+\rho_F=1$ is fulfilled. This phase is due to the fact that the boson-boson and fermion-fermion repulsive interaction parameters are greater than the boson-fermion. It favors that the bosons and fermions occupy different sites within the confinement potential, configuring the coexists of bosonic and fermionic  Mott insulator state. This result was reported experimentally in the same relationship $N_F/N_B=1$ for a system composed by the bosonic isotope of $^{174}Yb$ mixed with the fermionic isotope of $^{173}Yb$ considering $U_{BB}<\left| U_{BF}\right| <U_{FF}$\cite{Sugawa11}. However, because the fermionic isotope of $^{173}Yb$ have six nuclear spin components the boson-boson interaction to create a bosonic Mott insulator is low. The above indicates that the lower internal degree of freedom in fermions, greater the repulsive interaction between bosons must be, to generate a Mott insulator state as shown in the results of this investigation (figure \ref{insulator}) where the relation is $U_{BB}>U_{BF}<U_{FF}$.

When considering the same densities $\rho_B=\rho_F=1/3$ and $\rho_B=\rho_F=1/2$, with different relations between interaction parameters, we found that transmutation is also present and a new ground state arise as shown in figure \ref{separation}. For the case $\rho_B=\rho_F=1/3$, $U_{BB}=8, U_{BF}=9$ and $U_{FF}=7$ as shown in the figure \ref{separation}(a), where bosons (red squares) are more confined than fermions (black circles) in the ends of trap giving rise to a phase separation where bosons are in band insulator state and fermions are in a superfluid state as verified by boson (green diamonds) and fermion (blue triangles) compressibility. In the center trap fermions are repelled from it, while bosons are kept in a insulator state, giving rise to a phase separation where there are no fermions and bosons are in a insulator state as verified by boson (green diamonds) and fermion (blue triangles) compressibility. 

A similar ground state emerge where $\rho_B=\rho_F=1/2$, $U_{BB}=4, U_{BF}=8$ and $U_{FF}=12$ (figure \ref{separation}(b)), where the relationship $\rho_B+\rho_F=1$ is fulfilled. This distribution is due to the fact that  boson-fermion repulsive interaction parameter, is greater than boson-boson and the on-site fermion-fermion interaction is the largest, which favors that bosons occupy different sites within the confinement potential, configuring a insulator state \cite{Sugawa11, Wang12}.

\section{CONCLUSION}\label{sec:Conclusions}

In conclusion, using the DMRG we study the ground-state density distribution of a mixture of scalar bosons in the soft-core limit and spin one-half fermions in the framework of the one-dimensional harmonic trap. The most important result of our research, which has not been reported for these systems, is that there a Bose-Fermi transmutation  where the number occupation of the species in the ground state tends to be reversed, when Bose and Fermi densities are exchanged. 

The fact that $U_{BB}> 0$ induce a fermion-fermion coupling, indicates that there is a limit value of the boson-boson repulsive interaction parameter, from which the Bose-Fermi transmutation is generated. By varying the interaction between atoms, we find that when $U_{BB}>U_{BF}<U_{FF}$, the ground state in the center of trap is composed by the coexistence of bosonic and fermionic Mott plateau. It is important to highlight that the lower internal degree of freedom in fermions, greater the repulsive interaction between bosons must be, to generate a Mott insulator state. When considering the relative strength of the on-site interaction parameters $U_{BB}<U_{BF}<U_{FF}$, we find that fermions are repelled out of the center trap, while bosons occupy the central region. Phase separation of boson and fermion components is due to the fact that boson-fermion repulsive interaction parameter, is greater than boson-boson and the on-site fermion-fermion interaction is the largest, which favors that bosons occupy different sites within the confinement potential configuring a insulator state.  Our results are relevant for the ongoing experiment on ultracold mixtures of atomic gas 1D interacting Bose-Fermi mixture, with particular attention in the case of bosonic and fermionic isotopes of the same element, such as $^{170} Yb-^{171}Yb$.

\section{ACKNOWLEDGMENTS}

R.A. is thankful for the support of Departamento Administrativo de Ciencia, Tecnología e Innovación (COLCIENCIAS) (Grant No. FP44842-135-2017). 
R.A. thanks to Fundación universitaria los Libertadores for support during the completion of this work.

\bibliography{apssamp}% Produces the bibliography via BibTeX.

\providecommand{\noopsort}[1]{}\providecommand{\singleletter}[1]{#1}%
\begin{thebibliography}{84}%
\makeatletter
\providecommand \@ifxundefined [1]{%
 \@ifx{#1\undefined}
}%
\providecommand \@ifnum [1]{%
 \ifnum #1\expandafter \@firstoftwo
 \else \expandafter \@secondoftwo
 \fi
}%
\providecommand \@ifx [1]{%
 \ifx #1\expandafter \@firstoftwo
 \else \expandafter \@secondoftwo
 \fi
}%
\providecommand \natexlab [1]{#1}%
\providecommand \enquote  [1]{``#1''}%
\providecommand \bibnamefont  [1]{#1}%
\providecommand \bibfnamefont [1]{#1}%
\providecommand \citenamefont [1]{#1}%
\providecommand \href@noop [0]{\@secondoftwo}%
\providecommand \href [0]{\begingroup \@sanitize@url \@href}%
\providecommand \@href[1]{\@@startlink{#1}\@@href}%
\providecommand \@@href[1]{\endgroup#1\@@endlink}%
\providecommand \@sanitize@url [0]{\catcode `\\12\catcode `\$12\catcode
  `\&12\catcode `\#12\catcode `\^12\catcode `\_12\catcode `\%12\relax}%
\providecommand \@@startlink[1]{}%
\providecommand \@@endlink[0]{}%
\providecommand \url  [0]{\begingroup\@sanitize@url \@url }%
\providecommand \@url [1]{\endgroup\@href {#1}{\urlprefix }}%
\providecommand \urlprefix  [0]{URL }%
\providecommand \Eprint [0]{\href }%
\providecommand \doibase [0]{http://dx.doi.org/}%
\providecommand \selectlanguage [0]{\@gobble}%
\providecommand \bibinfo  [0]{\@secondoftwo}%
\providecommand \bibfield  [0]{\@secondoftwo}%
\providecommand \translation [1]{[#1]}%
\providecommand \BibitemOpen [0]{}%
\providecommand \bibitemStop [0]{}%
\providecommand \bibitemNoStop [0]{.\EOS\space}%
\providecommand \EOS [0]{\spacefactor3000\relax}%
\providecommand \BibitemShut  [1]{\csname bibitem#1\endcsname}%
\let\auto@bib@innerbib\@empty
%</preamble>
\bibitem [{\citenamefont {Kerman}\ \emph {et~al.}(2000)\citenamefont {Kerman},
  \citenamefont {Vuleti\ifmmode~\acute{c}\else \'{c}\fi{}}, \citenamefont
  {Chin},\ and\ \citenamefont {Chu}}]{Kerman00}%
  \BibitemOpen
  \bibfield  {author} {\bibinfo {author} {\bibfnamefont {A.~J.}\ \bibnamefont
  {Kerman}}, \bibinfo {author} {\bibfnamefont {V.}~\bibnamefont
  {Vuleti\ifmmode~\acute{c}\else \'{c}\fi{}}}, \bibinfo {author} {\bibfnamefont
  {C.}~\bibnamefont {Chin}}, \ and\ \bibinfo {author} {\bibfnamefont
  {S.}~\bibnamefont {Chu}},\ }\href@noop {} {\bibfield  {journal} {\bibinfo
  {journal} {Phys. Rev. Lett.}\ }\textbf {\bibinfo {volume} {84}},\ \bibinfo
  {pages} {439} (\bibinfo {year} {2000})}\BibitemShut {NoStop}%
\bibitem [{\citenamefont {Greiner}\ \emph {et~al.}(2001)\citenamefont
  {Greiner}, \citenamefont {Bloch}, \citenamefont {Mandel}, \citenamefont
  {H\"ansch},\ and\ \citenamefont {Esslinger}}]{Greiner01}%
  \BibitemOpen
  \bibfield  {author} {\bibinfo {author} {\bibfnamefont {M.}~\bibnamefont
  {Greiner}}, \bibinfo {author} {\bibfnamefont {I.}~\bibnamefont {Bloch}},
  \bibinfo {author} {\bibfnamefont {O.}~\bibnamefont {Mandel}}, \bibinfo
  {author} {\bibfnamefont {T.~W.}\ \bibnamefont {H\"ansch}}, \ and\ \bibinfo
  {author} {\bibfnamefont {T.}~\bibnamefont {Esslinger}},\ }\href@noop {}
  {\bibfield  {journal} {\bibinfo  {journal} {Phys. Rev. Lett.}\ }\textbf
  {\bibinfo {volume} {87}},\ \bibinfo {pages} {160405} (\bibinfo {year}
  {2001})}\BibitemShut {NoStop}%
\bibitem [{\citenamefont {Greiner}\ \emph {et~al.}(2002)\citenamefont
  {Greiner}, \citenamefont {Mandel}, \citenamefont {Esslinger}, \citenamefont
  {Hänsch},\ and\ \citenamefont {Bloch}}]{Greiner02}%
  \BibitemOpen
  \bibfield  {author} {\bibinfo {author} {\bibfnamefont {M.}~\bibnamefont
  {Greiner}}, \bibinfo {author} {\bibfnamefont {O.}~\bibnamefont {Mandel}},
  \bibinfo {author} {\bibfnamefont {T.}~\bibnamefont {Esslinger}}, \bibinfo
  {author} {\bibfnamefont {T.}~\bibnamefont {Hänsch}}, \ and\ \bibinfo
  {author} {\bibfnamefont {I.}~\bibnamefont {Bloch}},\ }\href@noop {}
  {\bibfield  {journal} {\bibinfo  {journal} {Nature}\ }\textbf {\bibinfo
  {volume} {415}},\ \bibinfo {pages} {39} (\bibinfo {year} {2002})}\BibitemShut
  {NoStop}%
\bibitem [{\citenamefont {Orzel}\ \emph {et~al.}(2001)\citenamefont {Orzel},
  \citenamefont {Tuchman}, \citenamefont {Fenselau}, \citenamefont {Yasuda},\
  and\ \citenamefont {Kasevich}}]{Orzel01}%
  \BibitemOpen
  \bibfield  {author} {\bibinfo {author} {\bibfnamefont {C.}~\bibnamefont
  {Orzel}}, \bibinfo {author} {\bibfnamefont {A.}~\bibnamefont {Tuchman}},
  \bibinfo {author} {\bibfnamefont {M.}~\bibnamefont {Fenselau}}, \bibinfo
  {author} {\bibfnamefont {M.}~\bibnamefont {Yasuda}}, \ and\ \bibinfo {author}
  {\bibfnamefont {M.}~\bibnamefont {Kasevich}},\ }\href@noop {} {\bibfield
  {journal} {\bibinfo  {journal} {Science}\ }\textbf {\bibinfo {volume}
  {291}},\ \bibinfo {pages} {2386} (\bibinfo {year} {2001})}\BibitemShut
  {NoStop}%
\bibitem [{\citenamefont {Anglin}\ and\ \citenamefont
  {Ketterle}(2002)}]{Anglin02}%
  \BibitemOpen
  \bibfield  {author} {\bibinfo {author} {\bibfnamefont {J.~R.}\ \bibnamefont
  {Anglin}}\ and\ \bibinfo {author} {\bibfnamefont {W.}~\bibnamefont
  {Ketterle}},\ }\href@noop {} {\bibfield  {journal} {\bibinfo  {journal}
  {Nature}\ }\textbf {\bibinfo {volume} {416}},\ \bibinfo {pages} {211}
  (\bibinfo {year} {2002})}\BibitemShut {NoStop}%
\bibitem [{\citenamefont {Ospelkaus}\ \emph
  {et~al.}(2006{\natexlab{a}})\citenamefont {Ospelkaus}, \citenamefont
  {Ospelkaus}, \citenamefont {Humbert}, \citenamefont {Sengstock},\ and\
  \citenamefont {Bongs}}]{Ospelkaus06}%
  \BibitemOpen
  \bibfield  {author} {\bibinfo {author} {\bibfnamefont {S.}~\bibnamefont
  {Ospelkaus}}, \bibinfo {author} {\bibfnamefont {C.}~\bibnamefont
  {Ospelkaus}}, \bibinfo {author} {\bibfnamefont {L.}~\bibnamefont {Humbert}},
  \bibinfo {author} {\bibfnamefont {K.}~\bibnamefont {Sengstock}}, \ and\
  \bibinfo {author} {\bibfnamefont {K.}~\bibnamefont {Bongs}},\ }\href@noop {}
  {\bibfield  {journal} {\bibinfo  {journal} {Phys. Rev. Lett.}\ }\textbf
  {\bibinfo {volume} {97}},\ \bibinfo {pages} {120403} (\bibinfo {year}
  {2006}{\natexlab{a}})}\BibitemShut {NoStop}%
\bibitem [{\citenamefont {Hadzibabic}\ \emph {et~al.}(2003)\citenamefont
  {Hadzibabic}, \citenamefont {Gupta}, \citenamefont {Stan}, \citenamefont
  {Schunck}, \citenamefont {Zwierlein}, \citenamefont {Dieckmann},\ and\
  \citenamefont {Ketterle}}]{Hadzibabic03}%
  \BibitemOpen
  \bibfield  {author} {\bibinfo {author} {\bibfnamefont {Z.}~\bibnamefont
  {Hadzibabic}}, \bibinfo {author} {\bibfnamefont {S.}~\bibnamefont {Gupta}},
  \bibinfo {author} {\bibfnamefont {C.~A.}\ \bibnamefont {Stan}}, \bibinfo
  {author} {\bibfnamefont {C.~H.}\ \bibnamefont {Schunck}}, \bibinfo {author}
  {\bibfnamefont {M.~W.}\ \bibnamefont {Zwierlein}}, \bibinfo {author}
  {\bibfnamefont {K.}~\bibnamefont {Dieckmann}}, \ and\ \bibinfo {author}
  {\bibfnamefont {W.}~\bibnamefont {Ketterle}},\ }\href@noop {} {\bibfield
  {journal} {\bibinfo  {journal} {Phys. Rev. Lett.}\ }\textbf {\bibinfo
  {volume} {91}},\ \bibinfo {pages} {160401} (\bibinfo {year}
  {2003})}\BibitemShut {NoStop}%
\bibitem [{\citenamefont {Inouye}\ \emph {et~al.}(2004)\citenamefont {Inouye},
  \citenamefont {Goldwin}, \citenamefont {Olsen}, \citenamefont {Ticknor},
  \citenamefont {Bohn},\ and\ \citenamefont {Jin}}]{Inouye04}%
  \BibitemOpen
  \bibfield  {author} {\bibinfo {author} {\bibfnamefont {S.}~\bibnamefont
  {Inouye}}, \bibinfo {author} {\bibfnamefont {J.}~\bibnamefont {Goldwin}},
  \bibinfo {author} {\bibfnamefont {M.~L.}\ \bibnamefont {Olsen}}, \bibinfo
  {author} {\bibfnamefont {C.}~\bibnamefont {Ticknor}}, \bibinfo {author}
  {\bibfnamefont {J.~L.}\ \bibnamefont {Bohn}}, \ and\ \bibinfo {author}
  {\bibfnamefont {D.~S.}\ \bibnamefont {Jin}},\ }\href@noop {} {\bibfield
  {journal} {\bibinfo  {journal} {Phys. Rev. Lett.}\ }\textbf {\bibinfo
  {volume} {93}},\ \bibinfo {pages} {183201} (\bibinfo {year}
  {2004})}\BibitemShut {NoStop}%
\bibitem [{\citenamefont {Jessen}\ \emph {et~al.}(2001)\citenamefont {Jessen},
  \citenamefont {Haycock}, \citenamefont {Klose},\ and\ \citenamefont
  {Smith}}]{Jessen01}%
  \BibitemOpen
  \bibfield  {author} {\bibinfo {author} {\bibfnamefont {P.~S.}\ \bibnamefont
  {Jessen}}, \bibinfo {author} {\bibfnamefont {D.~L.}\ \bibnamefont {Haycock}},
  \bibinfo {author} {\bibfnamefont {G.}~\bibnamefont {Klose}}, \ and\ \bibinfo
  {author} {\bibfnamefont {G.~A.}\ \bibnamefont {Smith}},\ }\href@noop {}
  {\bibfield  {journal} {\bibinfo  {journal} {Quantum Inf. Comput.}\ }\textbf
  {\bibinfo {volume} {1}},\ \bibinfo {pages} {20} (\bibinfo {year}
  {2001})}\BibitemShut {NoStop}%
\bibitem [{\citenamefont {Duan}\ \emph {et~al.}(2003)\citenamefont {Duan},
  \citenamefont {Demler},\ and\ \citenamefont {Lukin}}]{Duan03}%
  \BibitemOpen
  \bibfield  {author} {\bibinfo {author} {\bibfnamefont {L.-M.}\ \bibnamefont
  {Duan}}, \bibinfo {author} {\bibfnamefont {E.}~\bibnamefont {Demler}}, \ and\
  \bibinfo {author} {\bibfnamefont {M.~D.}\ \bibnamefont {Lukin}},\ }\href@noop
  {} {\bibfield  {journal} {\bibinfo  {journal} {Phys. Rev. Lett.}\ }\textbf
  {\bibinfo {volume} {91}},\ \bibinfo {pages} {090402} (\bibinfo {year}
  {2003})}\BibitemShut {NoStop}%
\bibitem [{\citenamefont {Brennen}\ \emph {et~al.}(2000)\citenamefont
  {Brennen}, \citenamefont {Deutsch},\ and\ \citenamefont
  {Jessen}}]{Deutsch00}%
  \BibitemOpen
  \bibfield  {author} {\bibinfo {author} {\bibfnamefont {G.~K.}\ \bibnamefont
  {Brennen}}, \bibinfo {author} {\bibfnamefont {I.~H.}\ \bibnamefont
  {Deutsch}}, \ and\ \bibinfo {author} {\bibfnamefont {P.~S.}\ \bibnamefont
  {Jessen}},\ }\href@noop {} {\bibfield  {journal} {\bibinfo  {journal} {Phys.
  Rev. A}\ }\textbf {\bibinfo {volume} {61}},\ \bibinfo {pages} {062309}
  (\bibinfo {year} {2000})}\BibitemShut {NoStop}%
\bibitem [{\citenamefont {Jaksch}\ \emph {et~al.}(1999)\citenamefont {Jaksch},
  \citenamefont {Briegel}, \citenamefont {Cirac}, \citenamefont {Gardiner},\
  and\ \citenamefont {Zoller}}]{Jaksch99}%
  \BibitemOpen
  \bibfield  {author} {\bibinfo {author} {\bibfnamefont {D.}~\bibnamefont
  {Jaksch}}, \bibinfo {author} {\bibfnamefont {H.-J.}\ \bibnamefont {Briegel}},
  \bibinfo {author} {\bibfnamefont {J.~I.}\ \bibnamefont {Cirac}}, \bibinfo
  {author} {\bibfnamefont {C.~W.}\ \bibnamefont {Gardiner}}, \ and\ \bibinfo
  {author} {\bibfnamefont {P.}~\bibnamefont {Zoller}},\ }\href@noop {}
  {\bibfield  {journal} {\bibinfo  {journal} {Phys. Rev. Lett.}\ }\textbf
  {\bibinfo {volume} {82}},\ \bibinfo {pages} {1975} (\bibinfo {year}
  {1999})}\BibitemShut {NoStop}%
\bibitem [{\citenamefont {Garc\'{\i}a-Ripoll}\ and\ \citenamefont
  {Cirac}(2003)}]{Garcia03}%
  \BibitemOpen
  \bibfield  {author} {\bibinfo {author} {\bibfnamefont {J.~J.}\ \bibnamefont
  {Garc\'{\i}a-Ripoll}}\ and\ \bibinfo {author} {\bibfnamefont {J.~I.}\
  \bibnamefont {Cirac}},\ }\href@noop {} {\bibfield  {journal} {\bibinfo
  {journal} {Phys. Rev. Lett.}\ }\textbf {\bibinfo {volume} {90}},\ \bibinfo
  {pages} {127902} (\bibinfo {year} {2003})}\BibitemShut {NoStop}%
\bibitem [{\citenamefont {Dorner}\ \emph {et~al.}(2003)\citenamefont {Dorner},
  \citenamefont {Fedichev}, \citenamefont {Jaksch}, \citenamefont
  {Lewenstein},\ and\ \citenamefont {Zoller}}]{Dorner03}%
  \BibitemOpen
  \bibfield  {author} {\bibinfo {author} {\bibfnamefont {U.}~\bibnamefont
  {Dorner}}, \bibinfo {author} {\bibfnamefont {P.}~\bibnamefont {Fedichev}},
  \bibinfo {author} {\bibfnamefont {D.}~\bibnamefont {Jaksch}}, \bibinfo
  {author} {\bibfnamefont {M.}~\bibnamefont {Lewenstein}}, \ and\ \bibinfo
  {author} {\bibfnamefont {P.}~\bibnamefont {Zoller}},\ }\href@noop {}
  {\bibfield  {journal} {\bibinfo  {journal} {Phys. Rev. Lett.}\ }\textbf
  {\bibinfo {volume} {91}},\ \bibinfo {pages} {073601} (\bibinfo {year}
  {2003})}\BibitemShut {NoStop}%
\bibitem [{\citenamefont {Pachos}\ and\ \citenamefont
  {Knight}(2003)}]{Pachos03}%
  \BibitemOpen
  \bibfield  {author} {\bibinfo {author} {\bibfnamefont {J.~K.}\ \bibnamefont
  {Pachos}}\ and\ \bibinfo {author} {\bibfnamefont {P.~L.}\ \bibnamefont
  {Knight}},\ }\href@noop {} {\bibfield  {journal} {\bibinfo  {journal} {Phys.
  Rev. Lett.}\ }\textbf {\bibinfo {volume} {91}},\ \bibinfo {pages} {107902}
  (\bibinfo {year} {2003})}\BibitemShut {NoStop}%
\bibitem [{\citenamefont {Sachdev}(2011)}]{Sachdev99}%
  \BibitemOpen
  \bibfield  {author} {\bibinfo {author} {\bibfnamefont {S.}~\bibnamefont
  {Sachdev}},\ }\href {\doibase 10.1017/CBO9780511973765} {\emph {\bibinfo
  {title} {Quantum Phase Transitions}}},\ \bibinfo {edition} {2nd}\ ed.\
  (\bibinfo  {publisher} {Cambridge University Press},\ \bibinfo {year}
  {2011})\BibitemShut {NoStop}%
\bibitem [{\citenamefont {aksch}\ \emph {et~al.}(1998)\citenamefont {aksch},
  \citenamefont {Bruder}, \citenamefont {Cirac}, \citenamefont {Gardiner},\
  and\ \citenamefont {Zoller}}]{Jaksch98}%
  \BibitemOpen
  \bibfield  {author} {\bibinfo {author} {\bibfnamefont {D.}~\bibnamefont
  {aksch}}, \bibinfo {author} {\bibfnamefont {C.}~\bibnamefont {Bruder}},
  \bibinfo {author} {\bibfnamefont {J.~I.}\ \bibnamefont {Cirac}}, \bibinfo
  {author} {\bibfnamefont {C.~W.}\ \bibnamefont {Gardiner}}, \ and\ \bibinfo
  {author} {\bibfnamefont {P.}~\bibnamefont {Zoller}},\ }\href@noop {}
  {\bibfield  {journal} {\bibinfo  {journal} {Phys. Rev. Lett.}\ }\textbf
  {\bibinfo {volume} {81}},\ \bibinfo {pages} {3108} (\bibinfo {year}
  {1998})}\BibitemShut {NoStop}%
\bibitem [{\citenamefont {van Oosten}\ \emph {et~al.}(2001)\citenamefont {van
  Oosten}, \citenamefont {van~der Straten},\ and\ \citenamefont
  {Stoof}}]{van01}%
  \BibitemOpen
  \bibfield  {author} {\bibinfo {author} {\bibfnamefont {D.}~\bibnamefont {van
  Oosten}}, \bibinfo {author} {\bibfnamefont {P.}~\bibnamefont {van~der
  Straten}}, \ and\ \bibinfo {author} {\bibfnamefont {H.~T.~C.}\ \bibnamefont
  {Stoof}},\ }\href@noop {} {\bibfield  {journal} {\bibinfo  {journal} {Phys.
  Rev. A}\ }\textbf {\bibinfo {volume} {63}},\ \bibinfo {pages} {053601}
  (\bibinfo {year} {2001})}\BibitemShut {NoStop}%
\bibitem [{\citenamefont {Ruostekoski}\ \emph {et~al.}(2002)\citenamefont
  {Ruostekoski}, \citenamefont {Dunne},\ and\ \citenamefont
  {Javanainen}}]{Ruostekoski02}%
  \BibitemOpen
  \bibfield  {author} {\bibinfo {author} {\bibfnamefont {J.}~\bibnamefont
  {Ruostekoski}}, \bibinfo {author} {\bibfnamefont {G.~V.}\ \bibnamefont
  {Dunne}}, \ and\ \bibinfo {author} {\bibfnamefont {J.}~\bibnamefont
  {Javanainen}},\ }\href@noop {} {\bibfield  {journal} {\bibinfo  {journal}
  {Phys. Rev. Lett.}\ }\textbf {\bibinfo {volume} {88}},\ \bibinfo {pages}
  {180401} (\bibinfo {year} {2002})}\BibitemShut {NoStop}%
\bibitem [{\citenamefont {Hofstetter}\ \emph {et~al.}(2002)\citenamefont
  {Hofstetter}, \citenamefont {Cirac}, \citenamefont {Zoller}, \citenamefont
  {Demler},\ and\ \citenamefont {Lukin}}]{Hofstetter02}%
  \BibitemOpen
  \bibfield  {author} {\bibinfo {author} {\bibfnamefont {W.}~\bibnamefont
  {Hofstetter}}, \bibinfo {author} {\bibfnamefont {J.~I.}\ \bibnamefont
  {Cirac}}, \bibinfo {author} {\bibfnamefont {P.}~\bibnamefont {Zoller}},
  \bibinfo {author} {\bibfnamefont {E.}~\bibnamefont {Demler}}, \ and\ \bibinfo
  {author} {\bibfnamefont {M.~D.}\ \bibnamefont {Lukin}},\ }\href@noop {}
  {\bibfield  {journal} {\bibinfo  {journal} {Phys. Rev. Lett.}\ }\textbf
  {\bibinfo {volume} {89}},\ \bibinfo {pages} {220407} (\bibinfo {year}
  {2002})}\BibitemShut {NoStop}%
\bibitem [{\citenamefont {Paredes}\ and\ \citenamefont
  {Cirac}(2003)}]{Paredes03}%
  \BibitemOpen
  \bibfield  {author} {\bibinfo {author} {\bibfnamefont {B.}~\bibnamefont
  {Paredes}}\ and\ \bibinfo {author} {\bibfnamefont {J.~I.}\ \bibnamefont
  {Cirac}},\ }\href@noop {} {\bibfield  {journal} {\bibinfo  {journal} {Phys.
  Rev. Lett.}\ }\textbf {\bibinfo {volume} {90}},\ \bibinfo {pages} {150402}
  (\bibinfo {year} {2003})}\BibitemShut {NoStop}%
\bibitem [{\citenamefont {Recati}\ \emph {et~al.}(2003)\citenamefont {Recati},
  \citenamefont {Fedichev}, \citenamefont {Zwerger},\ and\ \citenamefont
  {Zoller}}]{Recati03}%
  \BibitemOpen
  \bibfield  {author} {\bibinfo {author} {\bibfnamefont {A.}~\bibnamefont
  {Recati}}, \bibinfo {author} {\bibfnamefont {P.~O.}\ \bibnamefont
  {Fedichev}}, \bibinfo {author} {\bibfnamefont {W.}~\bibnamefont {Zwerger}}, \
  and\ \bibinfo {author} {\bibfnamefont {P.}~\bibnamefont {Zoller}},\
  }\href@noop {} {\bibfield  {journal} {\bibinfo  {journal} {Phys. Rev. Lett.}\
  }\textbf {\bibinfo {volume} {90}},\ \bibinfo {pages} {020401} (\bibinfo
  {year} {2003})}\BibitemShut {NoStop}%
\bibitem [{\citenamefont {Büchler}\ \emph {et~al.}(2003)\citenamefont
  {Büchler}, \citenamefont {Blatter},\ and\ \citenamefont
  {Zwerger}}]{Buchler03}%
  \BibitemOpen
  \bibfield  {author} {\bibinfo {author} {\bibfnamefont {H.~P.}\ \bibnamefont
  {Büchler}}, \bibinfo {author} {\bibfnamefont {G.}~\bibnamefont {Blatter}}, \
  and\ \bibinfo {author} {\bibfnamefont {W.}~\bibnamefont {Zwerger}},\
  }\href@noop {} {\bibfield  {journal} {\bibinfo  {journal} {Phys. Rev. Lett.}\
  }\textbf {\bibinfo {volume} {90}},\ \bibinfo {pages} {130401} (\bibinfo
  {year} {2003})}\BibitemShut {NoStop}%
\bibitem [{\citenamefont {Bijlsma}\ \emph {et~al.}(2000)\citenamefont
  {Bijlsma}, \citenamefont {Heringa},\ and\ \citenamefont {Stoof}}]{Bijlsma00}%
  \BibitemOpen
  \bibfield  {author} {\bibinfo {author} {\bibfnamefont {M.~J.}\ \bibnamefont
  {Bijlsma}}, \bibinfo {author} {\bibfnamefont {B.~A.}\ \bibnamefont
  {Heringa}}, \ and\ \bibinfo {author} {\bibfnamefont {H.~T.~C.}\ \bibnamefont
  {Stoof}},\ }\href@noop {} {\bibfield  {journal} {\bibinfo  {journal} {Phys.
  Rev. A}\ }\textbf {\bibinfo {volume} {61}},\ \bibinfo {pages} {053601}
  (\bibinfo {year} {2000})}\BibitemShut {NoStop}%
\bibitem [{\citenamefont {Heiselberg}\ \emph {et~al.}(2000)\citenamefont
  {Heiselberg}, \citenamefont {Pethick}, \citenamefont {Smith},\ and\
  \citenamefont {Viverit}}]{Heiselberg00}%
  \BibitemOpen
  \bibfield  {author} {\bibinfo {author} {\bibfnamefont {H.}~\bibnamefont
  {Heiselberg}}, \bibinfo {author} {\bibfnamefont {C.~J.}\ \bibnamefont
  {Pethick}}, \bibinfo {author} {\bibfnamefont {H.}~\bibnamefont {Smith}}, \
  and\ \bibinfo {author} {\bibfnamefont {L.}~\bibnamefont {Viverit}},\
  }\href@noop {} {\bibfield  {journal} {\bibinfo  {journal} {Phys. Rev. Lett.}\
  }\textbf {\bibinfo {volume} {85}},\ \bibinfo {pages} {2418} (\bibinfo {year}
  {2000})}\BibitemShut {NoStop}%
\bibitem [{\citenamefont {Viverit}\ and\ \citenamefont
  {Giorgini}(2002)}]{Viverit02}%
  \BibitemOpen
  \bibfield  {author} {\bibinfo {author} {\bibfnamefont {L.}~\bibnamefont
  {Viverit}}\ and\ \bibinfo {author} {\bibfnamefont {S.}~\bibnamefont
  {Giorgini}},\ }\href@noop {} {\bibfield  {journal} {\bibinfo  {journal}
  {Phys. Rev. A}\ }\textbf {\bibinfo {volume} {66}},\ \bibinfo {pages} {063604}
  (\bibinfo {year} {2002})}\BibitemShut {NoStop}%
\bibitem [{\citenamefont {Pollet}\ \emph {et~al.}(2008)\citenamefont {Pollet},
  \citenamefont {Kollath}, \citenamefont {Schollw\"ock},\ and\ \citenamefont
  {Troyer}}]{Pollet08}%
  \BibitemOpen
  \bibfield  {author} {\bibinfo {author} {\bibfnamefont {L.}~\bibnamefont
  {Pollet}}, \bibinfo {author} {\bibfnamefont {C.}~\bibnamefont {Kollath}},
  \bibinfo {author} {\bibfnamefont {U.}~\bibnamefont {Schollw\"ock}}, \ and\
  \bibinfo {author} {\bibfnamefont {M.}~\bibnamefont {Troyer}},\ }\href@noop {}
  {\bibfield  {journal} {\bibinfo  {journal} {Phys. Rev. A}\ }\textbf {\bibinfo
  {volume} {77}},\ \bibinfo {pages} {023608} (\bibinfo {year}
  {2008})}\BibitemShut {NoStop}%
\bibitem [{\citenamefont {McNamara}\ \emph {et~al.}(2006)\citenamefont
  {McNamara}, \citenamefont {Jeltes}, \citenamefont {Tychkov}, \citenamefont
  {Hogervorst},\ and\ \citenamefont {Vassen}}]{McNamara06}%
  \BibitemOpen
  \bibfield  {author} {\bibinfo {author} {\bibfnamefont {J.~M.}\ \bibnamefont
  {McNamara}}, \bibinfo {author} {\bibfnamefont {T.}~\bibnamefont {Jeltes}},
  \bibinfo {author} {\bibfnamefont {A.~S.}\ \bibnamefont {Tychkov}}, \bibinfo
  {author} {\bibfnamefont {W.}~\bibnamefont {Hogervorst}}, \ and\ \bibinfo
  {author} {\bibfnamefont {W.}~\bibnamefont {Vassen}},\ }\href@noop {}
  {\bibfield  {journal} {\bibinfo  {journal} {Phys. Rev. Lett.}\ }\textbf
  {\bibinfo {volume} {97}},\ \bibinfo {pages} {080404} (\bibinfo {year}
  {2006})}\BibitemShut {NoStop}%
\bibitem [{\citenamefont {Klempt}\ \emph {et~al.}(2008)\citenamefont {Klempt},
  \citenamefont {Henninger}, \citenamefont {Topic}, \citenamefont {Will},
  \citenamefont {Falke}, \citenamefont {Ertmer},\ and\ \citenamefont
  {Arlt}}]{Klempt08}%
  \BibitemOpen
  \bibfield  {author} {\bibinfo {author} {\bibfnamefont {C.}~\bibnamefont
  {Klempt}}, \bibinfo {author} {\bibfnamefont {T.}~\bibnamefont {Henninger}},
  \bibinfo {author} {\bibfnamefont {O.}~\bibnamefont {Topic}}, \bibinfo
  {author} {\bibfnamefont {J.}~\bibnamefont {Will}}, \bibinfo {author}
  {\bibfnamefont {S.}~\bibnamefont {Falke}}, \bibinfo {author} {\bibfnamefont
  {W.}~\bibnamefont {Ertmer}}, \ and\ \bibinfo {author} {\bibfnamefont
  {J.}~\bibnamefont {Arlt}},\ }\href@noop {} {\bibfield  {journal} {\bibinfo
  {journal} {Eur. Phys. J. D}\ }\textbf {\bibinfo {volume} {48}},\ \bibinfo
  {pages} {121} (\bibinfo {year} {2008})}\BibitemShut {NoStop}%
\bibitem [{\citenamefont {Karpiuk}\ \emph {et~al.}(2006)\citenamefont
  {Karpiuk}, \citenamefont {Brewczyk},\ and\ \citenamefont
  {Rzazewski}}]{Karpiuk06}%
  \BibitemOpen
  \bibfield  {author} {\bibinfo {author} {\bibfnamefont {T.}~\bibnamefont
  {Karpiuk}}, \bibinfo {author} {\bibfnamefont {M.}~\bibnamefont {Brewczyk}}, \
  and\ \bibinfo {author} {\bibfnamefont {K.}~\bibnamefont {Rzazewski}},\
  }\href@noop {} {\bibfield  {journal} {\bibinfo  {journal} {Phys. Rev. A}\
  }\textbf {\bibinfo {volume} {73}},\ \bibinfo {pages} {053602} (\bibinfo
  {year} {2006})}\BibitemShut {NoStop}%
\bibitem [{\citenamefont {Schneider}\ \emph {et~al.}(2008)\citenamefont
  {Schneider}, \citenamefont {Hackerm{\"u}ller}, \citenamefont {Will},
  \citenamefont {Best}, \citenamefont {Bloch}, \citenamefont {Costi},
  \citenamefont {Helmes}, \citenamefont {Rasch},\ and\ \citenamefont
  {Rosch}}]{Schneider08}%
  \BibitemOpen
  \bibfield  {author} {\bibinfo {author} {\bibfnamefont {U.}~\bibnamefont
  {Schneider}}, \bibinfo {author} {\bibfnamefont {L.}~\bibnamefont
  {Hackerm{\"u}ller}}, \bibinfo {author} {\bibfnamefont {S.}~\bibnamefont
  {Will}}, \bibinfo {author} {\bibfnamefont {T.}~\bibnamefont {Best}}, \bibinfo
  {author} {\bibfnamefont {I.}~\bibnamefont {Bloch}}, \bibinfo {author}
  {\bibfnamefont {T.~A.}\ \bibnamefont {Costi}}, \bibinfo {author}
  {\bibfnamefont {R.~W.}\ \bibnamefont {Helmes}}, \bibinfo {author}
  {\bibfnamefont {D.}~\bibnamefont {Rasch}}, \ and\ \bibinfo {author}
  {\bibfnamefont {A.}~\bibnamefont {Rosch}},\ }\href {\doibase
  10.1126/science.1165449} {\bibfield  {journal} {\bibinfo  {journal}
  {Science}\ }\textbf {\bibinfo {volume} {322}},\ \bibinfo {pages} {1520}
  (\bibinfo {year} {2008})}\BibitemShut {NoStop}%
\bibitem [{\citenamefont {Lous}\ \emph {et~al.}(2018)\citenamefont {Lous},
  \citenamefont {Fritsche}, \citenamefont {Jag}, \citenamefont {Lehmann},
  \citenamefont {Kirilov}, \citenamefont {Huang},\ and\ \citenamefont
  {Grimm}}]{Lous18}%
  \BibitemOpen
  \bibfield  {author} {\bibinfo {author} {\bibfnamefont {R.~S.}\ \bibnamefont
  {Lous}}, \bibinfo {author} {\bibfnamefont {I.}~\bibnamefont {Fritsche}},
  \bibinfo {author} {\bibfnamefont {M.}~\bibnamefont {Jag}}, \bibinfo {author}
  {\bibfnamefont {F.}~\bibnamefont {Lehmann}}, \bibinfo {author} {\bibfnamefont
  {E.}~\bibnamefont {Kirilov}}, \bibinfo {author} {\bibfnamefont
  {B.}~\bibnamefont {Huang}}, \ and\ \bibinfo {author} {\bibfnamefont
  {R.}~\bibnamefont {Grimm}},\ }\href@noop {} {\bibfield  {journal} {\bibinfo
  {journal} {Phys. Rev. Lett.}\ }\textbf {\bibinfo {volume} {120}},\ \bibinfo
  {pages} {243403} (\bibinfo {year} {2018})}\BibitemShut {NoStop}%
\bibitem [{\citenamefont {Ospelkaus}\ \emph
  {et~al.}(2006{\natexlab{b}})\citenamefont {Ospelkaus}, \citenamefont
  {Ospelkaus}, \citenamefont {Wille}, \citenamefont {Succo}, \citenamefont
  {Ernst}, \citenamefont {Sengstock},\ and\ \citenamefont {Bongs}}]{Wille06}%
  \BibitemOpen
  \bibfield  {author} {\bibinfo {author} {\bibfnamefont {S.}~\bibnamefont
  {Ospelkaus}}, \bibinfo {author} {\bibfnamefont {C.}~\bibnamefont
  {Ospelkaus}}, \bibinfo {author} {\bibfnamefont {O.}~\bibnamefont {Wille}},
  \bibinfo {author} {\bibfnamefont {M.}~\bibnamefont {Succo}}, \bibinfo
  {author} {\bibfnamefont {P.}~\bibnamefont {Ernst}}, \bibinfo {author}
  {\bibfnamefont {K.}~\bibnamefont {Sengstock}}, \ and\ \bibinfo {author}
  {\bibfnamefont {K.}~\bibnamefont {Bongs}},\ }\href@noop {} {\bibfield
  {journal} {\bibinfo  {journal} {Physical review letters}\ }\textbf {\bibinfo
  {volume} {96}},\ \bibinfo {pages} {180403} (\bibinfo {year}
  {2006}{\natexlab{b}})}\BibitemShut {NoStop}%
\bibitem [{\citenamefont {Sugawa}\ \emph {et~al.}(2011)\citenamefont {Sugawa},
  \citenamefont {Inaba}, \citenamefont {Taie}, \citenamefont {Yamazaki},
  \citenamefont {Yamashita},\ and\ \citenamefont {Takahashi}}]{Sugawa11}%
  \BibitemOpen
  \bibfield  {author} {\bibinfo {author} {\bibfnamefont {S.}~\bibnamefont
  {Sugawa}}, \bibinfo {author} {\bibfnamefont {K.}~\bibnamefont {Inaba}},
  \bibinfo {author} {\bibfnamefont {S.}~\bibnamefont {Taie}}, \bibinfo {author}
  {\bibfnamefont {R.}~\bibnamefont {Yamazaki}}, \bibinfo {author}
  {\bibfnamefont {M.}~\bibnamefont {Yamashita}}, \ and\ \bibinfo {author}
  {\bibfnamefont {Y.}~\bibnamefont {Takahashi}},\ }\href@noop {} {\bibfield
  {journal} {\bibinfo  {journal} {Nature Phys.}\ }\textbf {\bibinfo {volume}
  {7}},\ \bibinfo {pages} {642} (\bibinfo {year} {2011})}\BibitemShut {NoStop}%
\bibitem [{\citenamefont {Akdeniz}\ \emph {et~al.}(2002)\citenamefont
  {Akdeniz}, \citenamefont {Vignolo}, \citenamefont {Minguzzi},\ and\
  \citenamefont {Tosi}}]{Akdeniz02}%
  \BibitemOpen
  \bibfield  {author} {\bibinfo {author} {\bibfnamefont {Z.}~\bibnamefont
  {Akdeniz}}, \bibinfo {author} {\bibfnamefont {P.}~\bibnamefont {Vignolo}},
  \bibinfo {author} {\bibfnamefont {A.}~\bibnamefont {Minguzzi}}, \ and\
  \bibinfo {author} {\bibfnamefont {M.~P.}\ \bibnamefont {Tosi}},\ }\href@noop
  {} {\bibfield  {journal} {\bibinfo  {journal} {J. Phys. B: At. Mol. Opt.
  Phys.}\ }\textbf {\bibinfo {volume} {35}},\ \bibinfo {pages} {L105} (\bibinfo
  {year} {2002})}\BibitemShut {NoStop}%
\bibitem [{\citenamefont {Vichi}\ \emph {et~al.}(1998)\citenamefont {Vichi},
  \citenamefont {Inguscio}, \citenamefont {Stringari},\ and\ \citenamefont
  {Tino}}]{Vichi98}%
  \BibitemOpen
  \bibfield  {author} {\bibinfo {author} {\bibfnamefont {L.}~\bibnamefont
  {Vichi}}, \bibinfo {author} {\bibfnamefont {M.}~\bibnamefont {Inguscio}},
  \bibinfo {author} {\bibfnamefont {S.}~\bibnamefont {Stringari}}, \ and\
  \bibinfo {author} {\bibfnamefont {G.~M.}\ \bibnamefont {Tino}},\ }\href@noop
  {} {\bibfield  {journal} {\bibinfo  {journal} {J. Phys. B: At. Mol. Opt.
  Phys.}\ }\textbf {\bibinfo {volume} {31}},\ \bibinfo {pages} {L899} (\bibinfo
  {year} {1998})}\BibitemShut {NoStop}%
\bibitem [{\citenamefont {Roth}\ and\ \citenamefont
  {Feldmeier}(2002)}]{Roth02}%
  \BibitemOpen
  \bibfield  {author} {\bibinfo {author} {\bibfnamefont {R.}~\bibnamefont
  {Roth}}\ and\ \bibinfo {author} {\bibfnamefont {H.}~\bibnamefont
  {Feldmeier}},\ }\href@noop {} {\bibfield  {journal} {\bibinfo  {journal}
  {Phys. Rev. A}\ }\textbf {\bibinfo {volume} {65}},\ \bibinfo {pages} {021603}
  (\bibinfo {year} {2002})}\BibitemShut {NoStop}%
\bibitem [{\citenamefont {Liu}\ \emph {et~al.}(2003)\citenamefont {Liu},
  \citenamefont {Modugno},\ and\ \citenamefont {Hu}}]{Liu03}%
  \BibitemOpen
  \bibfield  {author} {\bibinfo {author} {\bibfnamefont {X.-J.}\ \bibnamefont
  {Liu}}, \bibinfo {author} {\bibfnamefont {M.}~\bibnamefont {Modugno}}, \ and\
  \bibinfo {author} {\bibfnamefont {H.}~\bibnamefont {Hu}},\ }\href@noop {}
  {\bibfield  {journal} {\bibinfo  {journal} {Phys. Rev. A}\ }\textbf {\bibinfo
  {volume} {68}},\ \bibinfo {pages} {053605} (\bibinfo {year}
  {2003})}\BibitemShut {NoStop}%
\bibitem [{\citenamefont {Modugno}\ \emph {et~al.}(2003)\citenamefont
  {Modugno}, \citenamefont {Ferlaino}, \citenamefont {Riboli}, \citenamefont
  {Roati}, \citenamefont {Modugno},\ and\ \citenamefont
  {Inguscio}}]{Modugno03}%
  \BibitemOpen
  \bibfield  {author} {\bibinfo {author} {\bibfnamefont {M.}~\bibnamefont
  {Modugno}}, \bibinfo {author} {\bibfnamefont {F.}~\bibnamefont {Ferlaino}},
  \bibinfo {author} {\bibfnamefont {F.}~\bibnamefont {Riboli}}, \bibinfo
  {author} {\bibfnamefont {G.}~\bibnamefont {Roati}}, \bibinfo {author}
  {\bibfnamefont {G.}~\bibnamefont {Modugno}}, \ and\ \bibinfo {author}
  {\bibfnamefont {M.}~\bibnamefont {Inguscio}},\ }\href@noop {} {\bibfield
  {journal} {\bibinfo  {journal} {Phys. Rev. A}\ }\textbf {\bibinfo {volume}
  {68}},\ \bibinfo {pages} {043626} (\bibinfo {year} {2003})}\BibitemShut
  {NoStop}%
\bibitem [{\citenamefont {Adhikari}(2004)}]{Adhikari04}%
  \BibitemOpen
  \bibfield  {author} {\bibinfo {author} {\bibfnamefont {S.~K.}\ \bibnamefont
  {Adhikari}},\ }\href@noop {} {\bibfield  {journal} {\bibinfo  {journal}
  {Phys. Rev. A}\ }\textbf {\bibinfo {volume} {70}},\ \bibinfo {pages} {043617}
  (\bibinfo {year} {2004})}\BibitemShut {NoStop}%
\bibitem [{\citenamefont {Miyakawa}\ \emph {et~al.}(2001)\citenamefont
  {Miyakawa}, \citenamefont {Suzuki},\ and\ \citenamefont {Yabu}}]{Miyakawa01}%
  \BibitemOpen
  \bibfield  {author} {\bibinfo {author} {\bibfnamefont {T.}~\bibnamefont
  {Miyakawa}}, \bibinfo {author} {\bibfnamefont {T.}~\bibnamefont {Suzuki}}, \
  and\ \bibinfo {author} {\bibfnamefont {H.}~\bibnamefont {Yabu}},\ }\href@noop
  {} {\bibfield  {journal} {\bibinfo  {journal} {Phys. Rev. A}\ }\textbf
  {\bibinfo {volume} {64}},\ \bibinfo {pages} {033611} (\bibinfo {year}
  {2001})}\BibitemShut {NoStop}%
\bibitem [{\citenamefont {Karpiuk}\ \emph {et~al.}(2005)\citenamefont
  {Karpiuk}, \citenamefont {Brewczyk}, \citenamefont {Gajda},\ and\
  \citenamefont {Rzazewski}}]{Karpiuk05}%
  \BibitemOpen
  \bibfield  {author} {\bibinfo {author} {\bibfnamefont {T.}~\bibnamefont
  {Karpiuk}}, \bibinfo {author} {\bibfnamefont {M.}~\bibnamefont {Brewczyk}},
  \bibinfo {author} {\bibfnamefont {M.}~\bibnamefont {Gajda}}, \ and\ \bibinfo
  {author} {\bibfnamefont {K.}~\bibnamefont {Rzazewski}},\ }\href@noop {}
  {\bibfield  {journal} {\bibinfo  {journal} {Journal of Physics B: Atomic,
  Molecular and Optical Physics}\ }\textbf {\bibinfo {volume} {38}},\ \bibinfo
  {pages} {L215} (\bibinfo {year} {2005})}\BibitemShut {NoStop}%
\bibitem [{\citenamefont {Röthel}(2007)}]{Pelster07}%
  \BibitemOpen
  \bibfield  {author} {\bibinfo {author} {\bibfnamefont {P.~A.}\ \bibnamefont
  {Röthel}, \bibfnamefont {S.}},\ }\href@noop {} {\bibfield  {journal}
  {\bibinfo  {journal} {The European Physical Journal B}\ }\textbf {\bibinfo
  {volume} {59}},\ \bibinfo {pages} {343} (\bibinfo {year} {2007})}\BibitemShut
  {NoStop}%
\bibitem [{\citenamefont {Best}\ \emph {et~al.}(2009)\citenamefont {Best},
  \citenamefont {Will}, \citenamefont {Schneider}, \citenamefont
  {Hackerm\"uller}, \citenamefont {van Oosten}, \citenamefont {Bloch},\ and\
  \citenamefont {L\"uhmann}}]{Best09}%
  \BibitemOpen
  \bibfield  {author} {\bibinfo {author} {\bibfnamefont {T.}~\bibnamefont
  {Best}}, \bibinfo {author} {\bibfnamefont {S.}~\bibnamefont {Will}}, \bibinfo
  {author} {\bibfnamefont {U.}~\bibnamefont {Schneider}}, \bibinfo {author}
  {\bibfnamefont {L.}~\bibnamefont {Hackerm\"uller}}, \bibinfo {author}
  {\bibfnamefont {D.}~\bibnamefont {van Oosten}}, \bibinfo {author}
  {\bibfnamefont {I.}~\bibnamefont {Bloch}}, \ and\ \bibinfo {author}
  {\bibfnamefont {D.-S.}\ \bibnamefont {L\"uhmann}},\ }\href@noop {} {\bibfield
   {journal} {\bibinfo  {journal} {Phys. Rev. Lett.}\ }\textbf {\bibinfo
  {volume} {102}},\ \bibinfo {pages} {030408} (\bibinfo {year}
  {2009})}\BibitemShut {NoStop}%
\bibitem [{\citenamefont {Albus}\ \emph {et~al.}(2003)\citenamefont {Albus},
  \citenamefont {Illuminati},\ and\ \citenamefont {Eisert}}]{Albus03}%
  \BibitemOpen
  \bibfield  {author} {\bibinfo {author} {\bibfnamefont {A.}~\bibnamefont
  {Albus}}, \bibinfo {author} {\bibfnamefont {F.}~\bibnamefont {Illuminati}}, \
  and\ \bibinfo {author} {\bibfnamefont {J.}~\bibnamefont {Eisert}},\
  }\href@noop {} {\bibfield  {journal} {\bibinfo  {journal} {Phys. Rev. A}\
  }\textbf {\bibinfo {volume} {68}},\ \bibinfo {pages} {023606} (\bibinfo
  {year} {2003})}\BibitemShut {NoStop}%
\bibitem [{\citenamefont {Polak}\ and\ \citenamefont
  {Kope\ifmmode~\acute{c}\else \'{c}\fi{}}(2010)}]{Polak10}%
  \BibitemOpen
  \bibfield  {author} {\bibinfo {author} {\bibfnamefont {T.~P.}\ \bibnamefont
  {Polak}}\ and\ \bibinfo {author} {\bibfnamefont {T.~K.}\ \bibnamefont
  {Kope\ifmmode~\acute{c}\else \'{c}\fi{}}},\ }\href@noop {} {\bibfield
  {journal} {\bibinfo  {journal} {Phys. Rev. A}\ }\textbf {\bibinfo {volume}
  {81}},\ \bibinfo {pages} {043612} (\bibinfo {year} {2010})}\BibitemShut
  {NoStop}%
\bibitem [{\citenamefont {Titvinidze}\ \emph {et~al.}(2008)\citenamefont
  {Titvinidze}, \citenamefont {Snoek},\ and\ \citenamefont
  {Hofstetter}}]{Titvinidze08}%
  \BibitemOpen
  \bibfield  {author} {\bibinfo {author} {\bibfnamefont {I.}~\bibnamefont
  {Titvinidze}}, \bibinfo {author} {\bibfnamefont {M.}~\bibnamefont {Snoek}}, \
  and\ \bibinfo {author} {\bibfnamefont {W.}~\bibnamefont {Hofstetter}},\
  }\href@noop {} {\bibfield  {journal} {\bibinfo  {journal} {Phys. Rev. Lett.}\
  }\textbf {\bibinfo {volume} {100}},\ \bibinfo {pages} {100401} (\bibinfo
  {year} {2008})}\BibitemShut {NoStop}%
\bibitem [{\citenamefont {Avella}\ \emph {et~al.}(2019)\citenamefont {Avella},
  \citenamefont {Mendoza-Arenas}, \citenamefont {Franco},\ and\ \citenamefont
  {Silva-Valencia}}]{Avella19}%
  \BibitemOpen
  \bibfield  {author} {\bibinfo {author} {\bibfnamefont {R.}~\bibnamefont
  {Avella}}, \bibinfo {author} {\bibfnamefont {J.~J.}\ \bibnamefont
  {Mendoza-Arenas}}, \bibinfo {author} {\bibfnamefont {R.}~\bibnamefont
  {Franco}}, \ and\ \bibinfo {author} {\bibfnamefont {J.}~\bibnamefont
  {Silva-Valencia}},\ }\href@noop {} {\bibfield  {journal} {\bibinfo  {journal}
  {Phys. Rev. A}\ }\textbf {\bibinfo {volume} {100}},\ \bibinfo {pages}
  {063620} (\bibinfo {year} {2019})}\BibitemShut {NoStop}%
\bibitem [{\citenamefont {Mering}\ and\ \citenamefont
  {Fleischhauer}(2008)}]{Mering08}%
  \BibitemOpen
  \bibfield  {author} {\bibinfo {author} {\bibfnamefont {A.}~\bibnamefont
  {Mering}}\ and\ \bibinfo {author} {\bibfnamefont {M.}~\bibnamefont
  {Fleischhauer}},\ }\href@noop {} {\bibfield  {journal} {\bibinfo  {journal}
  {Phys. Rev. A}\ }\textbf {\bibinfo {volume} {77}},\ \bibinfo {pages} {023601}
  (\bibinfo {year} {2008})}\BibitemShut {NoStop}%
\bibitem [{\citenamefont {Bukov}\ and\ \citenamefont {Pollet}(2014)}]{Bukov14}%
  \BibitemOpen
  \bibfield  {author} {\bibinfo {author} {\bibfnamefont {M.}~\bibnamefont
  {Bukov}}\ and\ \bibinfo {author} {\bibfnamefont {L.}~\bibnamefont {Pollet}},\
  }\href@noop {} {\bibfield  {journal} {\bibinfo  {journal} {Phys. Rev. B}\
  }\textbf {\bibinfo {volume} {89}},\ \bibinfo {pages} {094502} (\bibinfo
  {year} {2014})}\BibitemShut {NoStop}%
\bibitem [{\citenamefont {Fehrmann}\ \emph {et~al.}(2004)\citenamefont
  {Fehrmann}, \citenamefont {Baranov}, \citenamefont {Lewenstein},\ and\
  \citenamefont {Santos}}]{Fehrmann04}%
  \BibitemOpen
  \bibfield  {author} {\bibinfo {author} {\bibfnamefont {H.}~\bibnamefont
  {Fehrmann}}, \bibinfo {author} {\bibfnamefont {M.}~\bibnamefont {Baranov}},
  \bibinfo {author} {\bibfnamefont {M.}~\bibnamefont {Lewenstein}}, \ and\
  \bibinfo {author} {\bibfnamefont {L.}~\bibnamefont {Santos}},\ }\href@noop {}
  {\bibfield  {journal} {\bibinfo  {journal} {Opt. Express}\ }\textbf {\bibinfo
  {volume} {12}},\ \bibinfo {pages} {55} (\bibinfo {year} {2004})}\BibitemShut
  {NoStop}%
\bibitem [{\citenamefont {Avella}\ \emph {et~al.}(2020)\citenamefont {Avella},
  \citenamefont {Mendoza-Arenas}, \citenamefont {Franco},\ and\ \citenamefont
  {Silva-Valencia}}]{Avella20}%
  \BibitemOpen
  \bibfield  {author} {\bibinfo {author} {\bibfnamefont {R.}~\bibnamefont
  {Avella}}, \bibinfo {author} {\bibfnamefont {J.~J.}\ \bibnamefont
  {Mendoza-Arenas}}, \bibinfo {author} {\bibfnamefont {R.}~\bibnamefont
  {Franco}}, \ and\ \bibinfo {author} {\bibfnamefont {J.}~\bibnamefont
  {Silva-Valencia}},\ }\href@noop {} {\bibfield  {journal} {\bibinfo  {journal}
  {Phys. Rev. A}\ }\textbf {\bibinfo {volume} {102}},\ \bibinfo {pages}
  {033341} (\bibinfo {year} {2020})}\BibitemShut {NoStop}%
\bibitem [{\citenamefont {Bergeman}\ \emph {et~al.}(2003)\citenamefont
  {Bergeman}, \citenamefont {Moore},\ and\ \citenamefont
  {Olshanii}}]{Bergeman03}%
  \BibitemOpen
  \bibfield  {author} {\bibinfo {author} {\bibfnamefont {T.}~\bibnamefont
  {Bergeman}}, \bibinfo {author} {\bibfnamefont {M.~G.}\ \bibnamefont {Moore}},
  \ and\ \bibinfo {author} {\bibfnamefont {M.}~\bibnamefont {Olshanii}},\
  }\href@noop {} {\bibfield  {journal} {\bibinfo  {journal} {Phys. Rev. Lett.}\
  }\textbf {\bibinfo {volume} {91}},\ \bibinfo {pages} {163201} (\bibinfo
  {year} {2003})}\BibitemShut {NoStop}%
\bibitem [{\citenamefont {Laird}\ \emph {et~al.}(2017)\citenamefont {Laird},
  \citenamefont {Shi}, \citenamefont {Parish},\ and\ \citenamefont
  {Levinsen}}]{Laird17}%
  \BibitemOpen
  \bibfield  {author} {\bibinfo {author} {\bibfnamefont {E.~K.}\ \bibnamefont
  {Laird}}, \bibinfo {author} {\bibfnamefont {Z.-Y.}\ \bibnamefont {Shi}},
  \bibinfo {author} {\bibfnamefont {M.~M.}\ \bibnamefont {Parish}}, \ and\
  \bibinfo {author} {\bibfnamefont {J.}~\bibnamefont {Levinsen}},\ }\href@noop
  {} {\bibfield  {journal} {\bibinfo  {journal} {Phys. Rev. A}\ }\textbf
  {\bibinfo {volume} {96}},\ \bibinfo {pages} {032701} (\bibinfo {year}
  {2017})}\BibitemShut {NoStop}%
\bibitem [{\citenamefont {Noda}\ \emph {et~al.}(2012)\citenamefont {Noda},
  \citenamefont {Peters}, \citenamefont {Kawakami},\ and\ \citenamefont
  {Pruschke}}]{Noda2012}%
  \BibitemOpen
  \bibfield  {author} {\bibinfo {author} {\bibfnamefont {K.}~\bibnamefont
  {Noda}}, \bibinfo {author} {\bibfnamefont {R.}~\bibnamefont {Peters}},
  \bibinfo {author} {\bibfnamefont {N.}~\bibnamefont {Kawakami}}, \ and\
  \bibinfo {author} {\bibfnamefont {T.}~\bibnamefont {Pruschke}},\ }\href@noop
  {} {\bibfield  {journal} {\bibinfo  {journal} {Phys. Rev. A}\ }\textbf
  {\bibinfo {volume} {85}},\ \bibinfo {pages} {043628} (\bibinfo {year}
  {2012})}\BibitemShut {NoStop}%
\bibitem [{\citenamefont {St\"oferle}\ \emph {et~al.}(2004)\citenamefont
  {St\"oferle}, \citenamefont {Moritz}, \citenamefont {Schori}, \citenamefont
  {K\"ohl},\ and\ \citenamefont {Esslinger}}]{Moritz04}%
  \BibitemOpen
  \bibfield  {author} {\bibinfo {author} {\bibfnamefont {T.}~\bibnamefont
  {St\"oferle}}, \bibinfo {author} {\bibfnamefont {H.}~\bibnamefont {Moritz}},
  \bibinfo {author} {\bibfnamefont {C.}~\bibnamefont {Schori}}, \bibinfo
  {author} {\bibfnamefont {M.}~\bibnamefont {K\"ohl}}, \ and\ \bibinfo {author}
  {\bibfnamefont {T.}~\bibnamefont {Esslinger}},\ }\href@noop {} {\bibfield
  {journal} {\bibinfo  {journal} {Phys. Rev. Lett.}\ }\textbf {\bibinfo
  {volume} {92}},\ \bibinfo {pages} {130403} (\bibinfo {year}
  {2004})}\BibitemShut {NoStop}%
\bibitem [{\citenamefont {Wang}\ \emph {et~al.}(2012)\citenamefont {Wang},
  \citenamefont {Hao},\ and\ \citenamefont {Zhang}}]{Wang12}%
  \BibitemOpen
  \bibfield  {author} {\bibinfo {author} {\bibfnamefont {H.}~\bibnamefont
  {Wang}}, \bibinfo {author} {\bibfnamefont {Y.}~\bibnamefont {Hao}}, \ and\
  \bibinfo {author} {\bibfnamefont {Y.}~\bibnamefont {Zhang}},\ }\href@noop {}
  {\bibfield  {journal} {\bibinfo  {journal} {Phys. Rev. A}\ }\textbf {\bibinfo
  {volume} {85}},\ \bibinfo {pages} {053630} (\bibinfo {year}
  {2012})}\BibitemShut {NoStop}%
\bibitem [{\citenamefont {Paredes}\ \emph {et~al.}(2004)\citenamefont
  {Paredes}, \citenamefont {Widera}, \citenamefont {Murg}, \citenamefont
  {Mandel}, \citenamefont {Fölling}, \citenamefont {Cirac}, \citenamefont
  {Shlyapnikov},\ and\ \citenamefont {I}}]{Paredes04}%
  \BibitemOpen
  \bibfield  {author} {\bibinfo {author} {\bibfnamefont {B.}~\bibnamefont
  {Paredes}}, \bibinfo {author} {\bibfnamefont {A.}~\bibnamefont {Widera}},
  \bibinfo {author} {\bibfnamefont {V.}~\bibnamefont {Murg}}, \bibinfo {author}
  {\bibfnamefont {Q.}~\bibnamefont {Mandel}}, \bibinfo {author} {\bibfnamefont
  {S.}~\bibnamefont {Fölling}}, \bibinfo {author} {\bibfnamefont
  {I.}~\bibnamefont {Cirac}}, \bibinfo {author} {\bibfnamefont {H.~T.~W.}\
  \bibnamefont {Shlyapnikov}, \bibfnamefont {G~V}}, \ and\ \bibinfo {author}
  {\bibfnamefont {B.}~\bibnamefont {I}},\ }\href@noop {} {\bibfield  {journal}
  {\bibinfo  {journal} {Nature}\ }\textbf {\bibinfo {volume} {429}},\ \bibinfo
  {pages} {277} (\bibinfo {year} {2004})}\BibitemShut {NoStop}%
\bibitem [{\citenamefont {Kinoshita}\ \emph {et~al.}(2004)\citenamefont
  {Kinoshita}, \citenamefont {Wenger},\ and\ \citenamefont
  {Weiss}}]{Kinoshita04}%
  \BibitemOpen
  \bibfield  {author} {\bibinfo {author} {\bibfnamefont {T.}~\bibnamefont
  {Kinoshita}}, \bibinfo {author} {\bibfnamefont {T.}~\bibnamefont {Wenger}}, \
  and\ \bibinfo {author} {\bibfnamefont {D.~S.}\ \bibnamefont {Weiss}},\
  }\href@noop {} {\bibfield  {journal} {\bibinfo  {journal} {Science}\ }\textbf
  {\bibinfo {volume} {305}},\ \bibinfo {pages} {1125} (\bibinfo {year}
  {2004})}\BibitemShut {NoStop}%
\bibitem [{\citenamefont {Cazalilla}\ and\ \citenamefont
  {Ho}(2003)}]{Cazalilla03}%
  \BibitemOpen
  \bibfield  {author} {\bibinfo {author} {\bibfnamefont {M.~A.}\ \bibnamefont
  {Cazalilla}}\ and\ \bibinfo {author} {\bibfnamefont {A.~F.}\ \bibnamefont
  {Ho}},\ }\href@noop {} {\bibfield  {journal} {\bibinfo  {journal} {Phys. Rev.
  Lett.}\ }\textbf {\bibinfo {volume} {91}},\ \bibinfo {pages} {150403}
  (\bibinfo {year} {2003})}\BibitemShut {NoStop}%
\bibitem [{\citenamefont {Mathey}\ \emph {et~al.}(2004)\citenamefont {Mathey},
  \citenamefont {Wang}, \citenamefont {Hofstetter}, \citenamefont {Lukin},\
  and\ \citenamefont {Demler}}]{Mathey04}%
  \BibitemOpen
  \bibfield  {author} {\bibinfo {author} {\bibfnamefont {L.}~\bibnamefont
  {Mathey}}, \bibinfo {author} {\bibfnamefont {D.~W.}\ \bibnamefont {Wang}},
  \bibinfo {author} {\bibfnamefont {W.}~\bibnamefont {Hofstetter}}, \bibinfo
  {author} {\bibfnamefont {M.~D.}\ \bibnamefont {Lukin}}, \ and\ \bibinfo
  {author} {\bibfnamefont {E.}~\bibnamefont {Demler}},\ }\href@noop {}
  {\bibfield  {journal} {\bibinfo  {journal} {Phys. Rev. Lett.}\ }\textbf
  {\bibinfo {volume} {93}},\ \bibinfo {pages} {120404} (\bibinfo {year}
  {2004})}\BibitemShut {NoStop}%
\bibitem [{\citenamefont {Imambekov}\ and\ \citenamefont
  {Demler}(2006{\natexlab{a}})}]{Imambekov06}%
  \BibitemOpen
  \bibfield  {author} {\bibinfo {author} {\bibfnamefont {A.}~\bibnamefont
  {Imambekov}}\ and\ \bibinfo {author} {\bibfnamefont {E.}~\bibnamefont
  {Demler}},\ }\href@noop {} {\bibfield  {journal} {\bibinfo  {journal} {Annals
  of Physics}\ }\textbf {\bibinfo {volume} {321}},\ \bibinfo {pages} {2390}
  (\bibinfo {year} {2006}{\natexlab{a}})}\BibitemShut {NoStop}%
\bibitem [{\citenamefont {Imambekov}\ and\ \citenamefont
  {Demler}(2006{\natexlab{b}})}]{Imambekov06R}%
  \BibitemOpen
  \bibfield  {author} {\bibinfo {author} {\bibfnamefont {A.}~\bibnamefont
  {Imambekov}}\ and\ \bibinfo {author} {\bibfnamefont {E.}~\bibnamefont
  {Demler}},\ }\href@noop {} {\bibfield  {journal} {\bibinfo  {journal} {Phys.
  Rev. A}\ }\textbf {\bibinfo {volume} {73}},\ \bibinfo {pages} {021602(R)}
  (\bibinfo {year} {2006}{\natexlab{b}})}\BibitemShut {NoStop}%
\bibitem [{\citenamefont {Batchelor}\ \emph {et~al.}(2005)\citenamefont
  {Batchelor}, \citenamefont {Bortz}, \citenamefont {Guan},\ and\ \citenamefont
  {Oelkers}}]{Batchelor05}%
  \BibitemOpen
  \bibfield  {author} {\bibinfo {author} {\bibfnamefont {M.~T.}\ \bibnamefont
  {Batchelor}}, \bibinfo {author} {\bibfnamefont {M.}~\bibnamefont {Bortz}},
  \bibinfo {author} {\bibfnamefont {X.-W.}\ \bibnamefont {Guan}}, \ and\
  \bibinfo {author} {\bibfnamefont {N.}~\bibnamefont {Oelkers}},\ }\href@noop
  {} {\bibfield  {journal} {\bibinfo  {journal} {Phys. Rev. A.}\ }\textbf
  {\bibinfo {volume} {72}},\ \bibinfo {pages} {061603(R)} (\bibinfo {year}
  {2005})}\BibitemShut {NoStop}%
\bibitem [{\citenamefont {Titvinidze}\ \emph {et~al.}(2009)\citenamefont
  {Titvinidze}, \citenamefont {Snoek},\ and\ \citenamefont
  {Hofstetter}}]{Titvinidze09}%
  \BibitemOpen
  \bibfield  {author} {\bibinfo {author} {\bibfnamefont {I.}~\bibnamefont
  {Titvinidze}}, \bibinfo {author} {\bibfnamefont {M.}~\bibnamefont {Snoek}}, \
  and\ \bibinfo {author} {\bibfnamefont {W.}~\bibnamefont {Hofstetter}},\
  }\href@noop {} {\bibfield  {journal} {\bibinfo  {journal} {Phys. Rev. B}\
  }\textbf {\bibinfo {volume} {79}},\ \bibinfo {pages} {144506} (\bibinfo
  {year} {2009})}\BibitemShut {NoStop}%
\bibitem [{\citenamefont {Hao}(2011)}]{Hao11}%
  \BibitemOpen
  \bibfield  {author} {\bibinfo {author} {\bibfnamefont {Y.-J.}\ \bibnamefont
  {Hao}},\ }\href@noop {} {\bibfield  {journal} {\bibinfo  {journal} {Chinese
  Physics B}\ }\textbf {\bibinfo {volume} {20}},\ \bibinfo {pages} {060307}
  (\bibinfo {year} {2011})}\BibitemShut {NoStop}%
\bibitem [{\citenamefont {Xiangguo~Yin}\ and\ \citenamefont
  {Zhang}(2009)}]{Xiangguo09}%
  \BibitemOpen
  \bibfield  {author} {\bibinfo {author} {\bibfnamefont {S.~C.}\ \bibnamefont
  {Xiangguo~Yin}}\ and\ \bibinfo {author} {\bibfnamefont {Y.}~\bibnamefont
  {Zhang}},\ }\href@noop {} {\bibfield  {journal} {\bibinfo  {journal} {Phys.
  Rev. A}\ }\textbf {\bibinfo {volume} {79}},\ \bibinfo {pages} {053604}
  (\bibinfo {year} {2009})}\BibitemShut {NoStop}%
\bibitem [{\citenamefont {Girardeau}\ and\ \citenamefont
  {Minguzzi}(2006)}]{Girardeau06}%
  \BibitemOpen
  \bibfield  {author} {\bibinfo {author} {\bibfnamefont {M.~D.}\ \bibnamefont
  {Girardeau}}\ and\ \bibinfo {author} {\bibfnamefont {A.}~\bibnamefont
  {Minguzzi}},\ }\href@noop {} {\bibfield  {journal} {\bibinfo  {journal}
  {Phys. Rev. Lett.}\ }\textbf {\bibinfo {volume} {99}},\ \bibinfo {pages}
  {230402} (\bibinfo {year} {2006})}\BibitemShut {NoStop}%
\bibitem [{\citenamefont {Guan}\ \emph {et~al.}(2009)\citenamefont {Guan},
  \citenamefont {Chen}, \citenamefont {Wang},\ and\ \citenamefont
  {Ma}}]{Guan09}%
  \BibitemOpen
  \bibfield  {author} {\bibinfo {author} {\bibfnamefont {L.}~\bibnamefont
  {Guan}}, \bibinfo {author} {\bibfnamefont {S.}~\bibnamefont {Chen}}, \bibinfo
  {author} {\bibfnamefont {Y.}~\bibnamefont {Wang}}, \ and\ \bibinfo {author}
  {\bibfnamefont {Z.-Q.}\ \bibnamefont {Ma}},\ }\href@noop {} {\bibfield
  {journal} {\bibinfo  {journal} {Phys. Rev. Lett.}\ }\textbf {\bibinfo
  {volume} {102}},\ \bibinfo {pages} {160402} (\bibinfo {year}
  {2009})}\BibitemShut {NoStop}%
\bibitem [{\citenamefont {Ma}\ \emph {et~al.}(2009)\citenamefont {Ma},
  \citenamefont {Chen}, \citenamefont {Guan},\ and\ \citenamefont
  {Wang}}]{Wang09}%
  \BibitemOpen
  \bibfield  {author} {\bibinfo {author} {\bibfnamefont {Z.-Q.}\ \bibnamefont
  {Ma}}, \bibinfo {author} {\bibfnamefont {S.}~\bibnamefont {Chen}}, \bibinfo
  {author} {\bibfnamefont {L.}~\bibnamefont {Guan}}, \ and\ \bibinfo {author}
  {\bibfnamefont {Y.}~\bibnamefont {Wang}},\ }\href@noop {} {\bibfield
  {journal} {\bibinfo  {journal} {J. Phys. A: Math. Theor.}\ }\textbf {\bibinfo
  {volume} {42}},\ \bibinfo {pages} {385210} (\bibinfo {year}
  {2009})}\BibitemShut {NoStop}%
\bibitem [{\citenamefont {Yang}(2009)}]{Yang09}%
  \BibitemOpen
  \bibfield  {author} {\bibinfo {author} {\bibfnamefont {C.}~\bibnamefont
  {Yang}},\ }\href@noop {} {\bibfield  {journal} {\bibinfo  {journal} {Chinese
  Phys. Lett.}\ }\textbf {\bibinfo {volume} {26}},\ \bibinfo {pages} {120504}
  (\bibinfo {year} {2009})}\BibitemShut {NoStop}%
\bibitem [{\citenamefont {Fang}\ \emph {et~al.}(2011)\citenamefont {Fang},
  \citenamefont {Vignolo}, \citenamefont {Gattobigio}, \citenamefont
  {Miniatura},\ and\ \citenamefont {Minguzzi}}]{Fang11}%
  \BibitemOpen
  \bibfield  {author} {\bibinfo {author} {\bibfnamefont {B.}~\bibnamefont
  {Fang}}, \bibinfo {author} {\bibfnamefont {P.}~\bibnamefont {Vignolo}},
  \bibinfo {author} {\bibfnamefont {M.}~\bibnamefont {Gattobigio}}, \bibinfo
  {author} {\bibfnamefont {C.}~\bibnamefont {Miniatura}}, \ and\ \bibinfo
  {author} {\bibfnamefont {A.}~\bibnamefont {Minguzzi}},\ }\href@noop {}
  {\bibfield  {journal} {\bibinfo  {journal} {Phys. Rev. A}\ }\textbf {\bibinfo
  {volume} {84}},\ \bibinfo {pages} {023626} (\bibinfo {year}
  {2011})}\BibitemShut {NoStop}%
\bibitem [{\citenamefont {Cheon}\ and\ \citenamefont
  {Shigehara}(1999)}]{Cheon99}%
  \BibitemOpen
  \bibfield  {author} {\bibinfo {author} {\bibfnamefont {T.}~\bibnamefont
  {Cheon}}\ and\ \bibinfo {author} {\bibfnamefont {T.}~\bibnamefont
  {Shigehara}},\ }\href@noop {} {\bibfield  {journal} {\bibinfo  {journal}
  {Phys. Rev. Lett.}\ }\textbf {\bibinfo {volume} {82}},\ \bibinfo {pages}
  {2536} (\bibinfo {year} {1999})}\BibitemShut {NoStop}%
\bibitem [{\citenamefont {Girardeau}\ and\ \citenamefont
  {Olshanii}(2004)}]{Girardeau04}%
  \BibitemOpen
  \bibfield  {author} {\bibinfo {author} {\bibfnamefont {M.~D.}\ \bibnamefont
  {Girardeau}}\ and\ \bibinfo {author} {\bibfnamefont {M.}~\bibnamefont
  {Olshanii}},\ }\href@noop {} {\bibfield  {journal} {\bibinfo  {journal}
  {Phys. Rev. A}\ }\textbf {\bibinfo {volume} {70}},\ \bibinfo {pages} {023608}
  (\bibinfo {year} {2004})}\BibitemShut {NoStop}%
\bibitem [{\citenamefont {Deuretzbacher}\ \emph {et~al.}(2008)\citenamefont
  {Deuretzbacher}, \citenamefont {Fredenhagen}, \citenamefont {Becker},
  \citenamefont {Bongs}, \citenamefont {Sengstock},\ and\ \citenamefont
  {Pfannkuche}}]{PhysRevLett.100.160405}%
  \BibitemOpen
  \bibfield  {author} {\bibinfo {author} {\bibfnamefont {F.}~\bibnamefont
  {Deuretzbacher}}, \bibinfo {author} {\bibfnamefont {K.}~\bibnamefont
  {Fredenhagen}}, \bibinfo {author} {\bibfnamefont {D.}~\bibnamefont {Becker}},
  \bibinfo {author} {\bibfnamefont {K.}~\bibnamefont {Bongs}}, \bibinfo
  {author} {\bibfnamefont {K.}~\bibnamefont {Sengstock}}, \ and\ \bibinfo
  {author} {\bibfnamefont {D.}~\bibnamefont {Pfannkuche}},\ }\href@noop {}
  {\bibfield  {journal} {\bibinfo  {journal} {Phys. Rev. Lett.}\ }\textbf
  {\bibinfo {volume} {100}},\ \bibinfo {pages} {160405} (\bibinfo {year}
  {2008})}\BibitemShut {NoStop}%
\bibitem [{\citenamefont {Valiente}(2020)}]{Valiente20}%
  \BibitemOpen
  \bibfield  {author} {\bibinfo {author} {\bibfnamefont {M.}~\bibnamefont
  {Valiente}},\ }\href@noop {} {\bibfield  {journal} {\bibinfo  {journal}
  {Phys. Rev. A}\ }\textbf {\bibinfo {volume} {102}},\ \bibinfo {pages}
  {053304} (\bibinfo {year} {2020})}\BibitemShut {NoStop}%
\bibitem [{\citenamefont {White}(1993)}]{White93}%
  \BibitemOpen
  \bibfield  {author} {\bibinfo {author} {\bibfnamefont {S.~R.}\ \bibnamefont
  {White}},\ }\href@noop {} {\bibfield  {journal} {\bibinfo  {journal} {Phys.
  Rev. B}\ }\textbf {\bibinfo {volume} {48}},\ \bibinfo {pages} {10345}
  (\bibinfo {year} {1993})}\BibitemShut {NoStop}%
\bibitem [{\citenamefont {Pai}\ \emph {et~al.}(1996)\citenamefont {Pai},
  \citenamefont {Pandit}, \citenamefont {Krishnamurthy},\ and\ \citenamefont
  {Ramasesha}}]{Pai96}%
  \BibitemOpen
  \bibfield  {author} {\bibinfo {author} {\bibfnamefont {R.~V.}\ \bibnamefont
  {Pai}}, \bibinfo {author} {\bibfnamefont {R.}~\bibnamefont {Pandit}},
  \bibinfo {author} {\bibfnamefont {H.~R.}\ \bibnamefont {Krishnamurthy}}, \
  and\ \bibinfo {author} {\bibfnamefont {S.}~\bibnamefont {Ramasesha}},\
  }\href@noop {} {\bibfield  {journal} {\bibinfo  {journal} {Phys. Rev. Lett.}\
  }\textbf {\bibinfo {volume} {76}},\ \bibinfo {pages} {2937} (\bibinfo {year}
  {1996})}\BibitemShut {NoStop}%
\bibitem [{\citenamefont {Rossini}\ and\ \citenamefont
  {Fazio}(2012)}]{Rossini12}%
  \BibitemOpen
  \bibfield  {author} {\bibinfo {author} {\bibfnamefont {D.}~\bibnamefont
  {Rossini}}\ and\ \bibinfo {author} {\bibfnamefont {R.}~\bibnamefont
  {Fazio}},\ }\href@noop {} {\bibfield  {journal} {\bibinfo  {journal} {New J.
  Phys.}\ }\textbf {\bibinfo {volume} {14}},\ \bibinfo {pages} {065012}
  (\bibinfo {year} {2012})}\BibitemShut {NoStop}%
\bibitem [{\citenamefont {Batrouni}\ \emph {et~al.}(2002)\citenamefont
  {Batrouni}, \citenamefont {Rousseau}, \citenamefont {Scalettar},
  \citenamefont {Rigol}, \citenamefont {Muramatsu}, \citenamefont {Denteneer},\
  and\ \citenamefont {Troyer}}]{Batrouni02}%
  \BibitemOpen
  \bibfield  {author} {\bibinfo {author} {\bibfnamefont {G.~G.}\ \bibnamefont
  {Batrouni}}, \bibinfo {author} {\bibfnamefont {V.}~\bibnamefont {Rousseau}},
  \bibinfo {author} {\bibfnamefont {R.~T.}\ \bibnamefont {Scalettar}}, \bibinfo
  {author} {\bibfnamefont {M.}~\bibnamefont {Rigol}}, \bibinfo {author}
  {\bibfnamefont {A.}~\bibnamefont {Muramatsu}}, \bibinfo {author}
  {\bibfnamefont {P.~J.~H.}\ \bibnamefont {Denteneer}}, \ and\ \bibinfo
  {author} {\bibfnamefont {M.}~\bibnamefont {Troyer}},\ }\href@noop {}
  {\bibfield  {journal} {\bibinfo  {journal} {Phys. Rev. Lett.}\ }\textbf
  {\bibinfo {volume} {89}},\ \bibinfo {pages} {117203} (\bibinfo {year}
  {2002})}\BibitemShut {NoStop}%
\bibitem [{\citenamefont {Rigol}\ and\ \citenamefont
  {Muramatsu}(2004)}]{Rigol08}%
  \BibitemOpen
  \bibfield  {author} {\bibinfo {author} {\bibfnamefont {M.}~\bibnamefont
  {Rigol}}\ and\ \bibinfo {author} {\bibfnamefont {A.}~\bibnamefont
  {Muramatsu}},\ }\href@noop {} {\bibfield  {journal} {\bibinfo  {journal}
  {Phys. Rev. A}\ }\textbf {\bibinfo {volume} {69}},\ \bibinfo {pages} {053612}
  (\bibinfo {year} {2004})}\BibitemShut {NoStop}%
\bibitem [{\citenamefont {Legeza}\ \emph {et~al.}(2003)\citenamefont {Legeza},
  \citenamefont {R\"oder},\ and\ \citenamefont {Hess}}]{Legeza03}%
  \BibitemOpen
  \bibfield  {author} {\bibinfo {author} {\bibfnamefont {O.}~\bibnamefont
  {Legeza}}, \bibinfo {author} {\bibfnamefont {J.}~\bibnamefont {R\"oder}}, \
  and\ \bibinfo {author} {\bibfnamefont {B.~A.}\ \bibnamefont {Hess}},\
  }\href@noop {} {\bibfield  {journal} {\bibinfo  {journal} {Phys. Rev. B.}\
  }\textbf {\bibinfo {volume} {67}},\ \bibinfo {pages} {125114} (\bibinfo
  {year} {2003})}\BibitemShut {NoStop}%
\bibitem [{\citenamefont {Dehkharghani}\ \emph {et~al.}(2017)\citenamefont
  {Dehkharghani}, \citenamefont {Bellotti},\ and\ \citenamefont
  {Zinner}}]{Dehkharghani17}%
  \BibitemOpen
  \bibfield  {author} {\bibinfo {author} {\bibfnamefont {A.~S.}\ \bibnamefont
  {Dehkharghani}}, \bibinfo {author} {\bibfnamefont {F.~F.}\ \bibnamefont
  {Bellotti}}, \ and\ \bibinfo {author} {\bibfnamefont {N.~T.}\ \bibnamefont
  {Zinner}},\ }\href@noop {} {\bibfield  {journal} {\bibinfo  {journal}
  {Journal of Physics B}\ }\textbf {\bibinfo {volume} {50}},\ \bibinfo {pages}
  {144002} (\bibinfo {year} {2017})}\BibitemShut {NoStop}%
\bibitem [{\citenamefont {Xianlong}(2013)}]{Xianlong13}%
  \BibitemOpen
  \bibfield  {author} {\bibinfo {author} {\bibfnamefont {G.}~\bibnamefont
  {Xianlong}},\ }\href@noop {} {\bibfield  {journal} {\bibinfo  {journal}
  {Physical Review A}\ }\textbf {\bibinfo {volume} {87}},\ \bibinfo {pages}
  {023628} (\bibinfo {year} {2013})}\BibitemShut {NoStop}%
\end{thebibliography}%

\end{document}